\documentclass[aps,prd,preprint,showpacs,floats]{revtex4}
\usepackage[dvips]{graphicx}
\usepackage{bm}
\usepackage{euscript,amsmath}
\def\calAslash{\rlap{\hspace{0.08cm}/}{{\EuScript A}}}
\def\nbarslash{\rlap{\hspace{0.02cm}/}{\bar n}}
\def\nslash{\rlap{\hspace{0.02cm}/} {n}}
\def\polslash{\rlap{\hspace{0.02cm}/} {\varepsilon}}
\begin{document}
\vspace*{-1cm}
\begin{flushright}
DESY 06-002 \\
January 2006\\
\end{flushright}
\title{\boldmath{$B \to K^\ast \ell^+ \ell^-$} {\Large \bf decay in soft-collinear effective theory }}
 \author{A. Ali}
 \email[E-mail address: ]{ahmed.ali@desy.de}
 \affiliation{Theory Group, Deutsches Elektronen-Synchrotron DESY, \\
  Notkestrasse 85, 22603 Hamburg, Germany.}
 \author{G. Kramer}
  \email[E-mail address: ]{gustav.kramer@desy.de}
  \author{Guohuai Zhu}
  \email[E-mail address: ]{guohuai.zhu@desy.de}
  \thanks{Alexander-von-Humboldt Fellow}
 \affiliation{ II. Institut f{\"u}r Theoretische Physik, Universit{\"a}t Hamburg, \\
Luruper Chaussee 149, 22761 Hamburg, Germany}
  \date{\today}
\begin{abstract}
We study the rare B decay $B \to K^\ast
\ell^+ \ell^-$ using soft-collinear effective theory (SCET).
At leading power in $1/m_b$, a factorization formula is obtained
 valid to all orders in $\alpha_s$. For phenomenological application,
we calculate the decay amplitude including order $\alpha_s$ corrections, and resum
the logarithms by evolving the matching coefficients from the hard
scale ${\cal O}(m_b)$ down to the scale $\sqrt{m_b \Lambda_h}$.
The branching ratio for $B \to K^\ast
\ell^+ \ell^-$ is uncertain due to the imprecise knowledge of the
soft form factors $\zeta_\perp (q^2)$ and $\zeta_\parallel (q^2)$.
Constraining the soft form factor $\zeta_\perp (q^2=0)$ from data on
$B \to K^* \gamma$ yields $\zeta_\perp (q^2=0)=0.32 \pm 0.02$. Using this input,
together with the light-cone sum rules to determine the $q^2$-dependence
of $\zeta_\perp (q^2)$ and the other soft form factor $\zeta_\parallel (q^2)$,
 we estimate
the partially integrated branching ratio in the range
$1~\mbox{GeV}^2 \le q^2 \le 7~\mbox{GeV}^2$ to be $(2.92^{+0.67}_{-0.61})
\times 10^{-7}$. We discuss how to reduce the form factor related uncertainty
by combining data on $B \to \rho (\to \pi \pi) \ell \nu_\ell$ and
$B\to K^* (\to K\pi) \ell^+\ell^-$.
 The forward-backward asymmetry is less sensitive
to the input parameters. In particular,
for the zero-point of
the forward-backward asymmetry in the standard model, we get
$q_0^2=(4.07^{+0.16}_{-0.13})
~\mbox{GeV}^2$. The scale dependence of $q_0^2$ is discussed in detail.
\end{abstract}
 \pacs{13.25.Hw, 12.39.St, 12.38.Bx}
\maketitle

\section{introduction}
The electroweak penguin decay $B \to K^\ast \ell^+ \ell^-$ is
loop-suppressed in the Standard Model(SM). It may therefore provide
a rigorous test of the SM and also put strong constraints on the
flavor physics beyond the SM.

Though the inclusive decay $B \to X_s \ell^+ \ell^-$ is better
understood theoretically using the Operator Product Expansion,
and the first direct experimental measurements of the dilepton
invariant mass spectrum and $m_X$-distribution are already at
 hand~\cite{Kaneko:2002mr,Aubert:2004it},
being an inclusive process, it is
extremely difficult to be measured in a hadron machine, such as the LHC,
which is the only collider, except for a Super-B factory,
that could provide enough luminosity for the precise study of the
decay distributions of such a rare process. In contrast, for the
exclusive decay $B \to K^\ast \ell^+ \ell^-$, the difficulty lies in
the imprecise knowledge of the underlying hadron dynamics.
 Experimentally, BaBar \cite{BaBar} and Belle \cite{Belle}
Collaborations have observed this rare decay with the branching ratios:
\begin{equation}
{\cal B}(B \to K^\ast \ell^+ \ell^- )=\left \{ \begin{array}{ll}
(7.8 ^{+1.9}_{-1.7} \pm 1.2) \times 10^{-7} & \mbox{(BaBar)}~,\\
(16.5^{+2.3}_{-2.2}\pm 0.9 \pm 0.4) \times 10^{-7} & \mbox{(Belle)}~.
\end{array} \right.
\end{equation}
We note that the Belle measurements are approximately a factor 2 higher
than the corresponding BaBar measurements.
In addition, Belle has published the measurements~\cite{Belle,Abe:2005km} of the so-called
forward-backward asymmetry (FBA)~\cite{Ali:1991is}. In particular, the best-fit
results by Belle for the Wilson coefficient ratios for negative value of $A_7$,
\begin{eqnarray}
\frac{A_9}{A_7} &=&-15.3 ^{+3.4}_{-4.8} \pm 1.1 , \nonumber\\
\frac{A_{10}}{A_7} &=&10.3 ^{+5.2}_{-3.5} \pm 1.8~,
\end{eqnarray}
are consistent with the SM values $A_9/A_7 \simeq -13.7$ and
$A_{10}/A_{7} \simeq +14.9$, evaluated in the NLO approximation
(see Table I).
With more data accumulated at the current B factories, and especially the huge
data that will be produced at the LHC, it is foreseeable that the dilepton
invariant mass spectrum and the FBA in this channel will be measured precisely in several
years from now, allowing a few \% measurements of the Wilson coefficient ratios and the
sign of $A_7$.

Theoretically, the exclusive decay $B \to K^\ast \ell^+ \ell^-$  has
been studied in a number of papers, see for example
\cite{Deshpande,Lim,Ali92,Greub94,Melikhov:1997zu,Ali00}. From the viewpoint of hadron
dynamics, the application of the QCD factorization approach
\cite{BBNS} to this channel \cite{Beneke01} deserves special mention, as we
shall be comparing our phenomenological analysis with the results obtained
in this paper.
 The emergence of an effective theory, called
soft-collinear effective theory (SCET)
\cite{Bauer1,Bauer2,Bauer3,Beneke02,Neubert02}, provides a
systematic and rigorous way to deal with the perturbative strong interaction
effects in  B decays in the heavy-quark expansion. A lot of theoretical work
has been done in SCET  related to the so-called heavy-to-light
transitions in $B$ decays, in particular, a demonstration of the soft-collinear
factorization~\cite{Bauer:2001yt,Neubert04B,Beneke04,Yang04}, a complete
catalogue of the
various 2-body and 3-body current
operators~\cite{Neubert02,Pirjol:2002km,Beneke04},
and the extension of SCET to two effective theories SCET$_I$ and SCET$_{II}$,
 with the two-step
matching QCD $\to$ SCET$_I$ $\to$  SCET$_{II}$~\cite{Bauer:2002aj}.
Among various phenomenological applications reported in the literature,
 SCET has been used to prove the factorization of radiative
$B \to V \gamma$ decays at leading power in $1/m_b$ and to all
orders in $\alpha_s$~\cite{Neubert05,Kim03}. Likewise, SCET, in combination
with the heavy-hadron chiral perturbation theory, has also been used to
study the forward-backward asymmetry in the non-resonant decay
$B \to K \pi \ell^+ \ell^- $ in certain kinematic
 region~\cite{Grinstein:2005ud}.
In this paper, our aim
is to use SCET in the decay $B \to K^\ast \ell^+ \ell^-$. Due to the
similarity between $B \to K^\ast \gamma$ and $B \to K^\ast \ell^+
\ell^-$ decays, our approach is quite similar to the earlier
SCET-based studies~\cite{Neubert05,Kim03}, in particular to
Ref.~\cite{Neubert05}. Moreover, an analysis of the exclusive radiative
and semileptonic decays $ B \to K^* \gamma$ and $B \to K^* \ell^+ \ell^-$ in
SCET can be combined with data to reduce the
uncertainties in the input parameters. In particular, as we show here, the
location of the forward-backward asymmetry in $B \to K^* \ell^+\ell^-$ can be
predicted more precisely than is the case in the existing literature.

It is well known that, when $q^2$, the momentum squared of the lepton
pair, is comparable to  $M^2_{J/\psi}$, the resonant
charmonium contributions become very important, for which there is
no model-independent treatment yet. Likewise, for higher $q^2$-values,
higher $\psi$-resonances $(\psi^\prime, \psi^{\prime\prime},...)$ have to be
included. Thus, in the following we will
restrict ourselves to the region $1~\mbox{GeV}^2 <q^2 < 7
~\mbox{GeV}^2$, which is dominated by the short-distance contribution.
 Note that the lower
cut-off $1~\mbox{GeV}^2$ is taken here because, as we shall see later,
when $q^2$ is very small, say $q^2 \sim {\cal O}(\Lambda_{QCD}^2)$,
the factorization of the annihilation topology breaks down.
In this kinematic region, a
factorization formula for the decay amplitude of $B \to K^* \ell^+\ell^-$,
which holds to
 $O(\alpha_s)$ at the leading power in $1/m_b$,
has been derived in Ref.~\cite{Beneke01} using the QCD factorization approach
 We shall
derive the factorization of the decay amplitude of  $B \to K^* \ell^+\ell^-$
in SCET, which {\it formally} coincides with the formula obtained  
by Beneke et al.~\cite{Beneke01}, but is valid to all orders of $\alpha_s$: 
\begin{equation}
\langle K_a^\ast \ell^+ \ell^- \vert H_{eff} \vert B \rangle = T^I_a(q^2)
\zeta_a(q^2) + \sum_{\pm} \int_0^\infty \frac{d\omega}{\omega}
\phi^{B}_{\pm}(\omega) \int_0^1 du ~\phi_{K^\ast}^{ a}(u)T^{II}_{a,\pm}
(\omega, u,q^2)~,
\end{equation}
where $a=\parallel,\perp$ denotes the polarization of the $K^\ast$
meson. The functions $T^I$ and $T^{II}$ are perturbatively
calculable. $\zeta_a(q^2)$ are the soft form factors defined in SCET while
$\phi^{B}_{\pm}$ and $\phi_{K^\ast}^{a}$ are the light-cone distribution
amplitudes (LCDAs) for the B and $K^\ast$ mesons, respectively.
Compared to the earlier results of Ref.
\cite{Beneke01}, obtained in the QCD factorization approach,
the main phenomenological improvement is that for
the hard scattering function $T^{II}$, the perturbative logarithms
are summed from the hard scale $\mu_b \sim {\cal O}(m_b)$ down to
the intermediate scale $\mu_\ell\equiv \sqrt{\mu_b \Lambda_h}$, where $\Lambda_h$
represents a typical hadronic scale. Note also that the
definitions of the soft form factors $\zeta_a(q^2)$ for our SCET currents,
defined subsequently in section 2, are different
from those of Ref.~\cite{Beneke01}, a point to which we will return later in
section 3. Hence, 
the explicit expressions for $T^I$, derived here and in   Ref.~\cite{Beneke01}
are also different.

This paper is organized as follows: In section II, we
briefly review the basic ideas and notations of SCET. We then list the relevant
effective operators in SCET and do the explicit matching calculations
from QCD to SCET$_I$~(Sec. II A) and from SCET$_I$ to SCET$_{II}$~(Sec.~II B).
The matrix elements of the effective SCET operators are given in Sec.~II C.
At the end of this section, the logarithmic resummation in SCET$_I$ is discussed.
In section III, we consider some phenomenological aspects of the $B \to K^\ast \ell^+
\ell^-$ decay. We first specify the input parameters, especially the soft form
factors $\zeta_{\perp,\parallel}(q^2)$~(Sec. III A), which are the cause of the
largest theoretical uncertainty. We use the $q^2$-dependence of the related
QCD form factors in the LC-QCD sum rule approach, but fix the
normalization of these soft form factors using constraints from data on the
exclusive decays $B \to K^* \gamma$.
In Sec. III B, we work out numerically
the evolution of the B-type SCET$_I$ matching coefficients, defined earlier in Sec.~II. We then give the dilepton
invariant mass spectrum and the forward-backward asymmetry in the decay
$B \to K^\ast \ell^+ \ell^-$, and compare the integrated branching ratios with the measurements from
BaBar and Belle (Sec.~III C). We end with a summary of our results in
section IV and suggestions for future measurements to reduce the model
dependence due to the form factors and other input parameters.

\section{SCET analysis of $B \to K^\ast \ell^+ \ell^-$}

For the $b \to s$ transitions, the weak effective Hamiltonian can be written as
\begin{equation}\label{Heff}
H_{eff}=-\frac{G_F}{\sqrt{2}}V_{ts}^\ast V_{tb} \sum_{i=1}^{10} C_i
(\mu) Q_i (\mu)~,
\end{equation}
where we have neglected the contribution proportional to $V_{us}^\ast V_{ub}$
in the penguin (loop) amplitudes,
which is doubly Cabibbo-suppressed, and have used the unitarity of the
CKM matrix to factorize the overall CKM-matrix element dependence. We use the operator basis introduced in
\cite{Misiak97,Beneke01}:
\begin{equation}
\begin{aligned}
Q_1&= (\bar{s} T^A c )_{V-A}
         ( \bar{c} T^A b )_{V-A}~,\hspace*{1.3cm}
Q_2= (\bar{s} c)_{V-A}
         ( \bar{c} b )_{V-A}~,\\
Q_3&= 2~(\bar{s} b)_{V-A}
      \sum \limits_{q}
     ( \bar{q} \gamma^\mu q )~,\hspace*{1.6cm}
Q_4= 2~(\bar{s} T^A b)_{V-A}
      \sum \limits_{q}
     ( \bar{q} \gamma^\mu T^A q )~,\\
Q_5&=2~\bar{s}\gamma_\mu \gamma_\nu \gamma_\rho (1-\gamma_5)b
     ~\sum \limits_q (\bar{q}\gamma^\mu \gamma^\nu \gamma^\rho q
     )~,\\
 Q_6&=2~\bar{s}\gamma_\mu \gamma_\nu \gamma_\rho
     (1-\gamma_5)T^A b~\sum \limits_q
     (\bar{q}\gamma^\mu \gamma^\nu \gamma^\rho T^A q )~,\\
Q_7&=-\frac{g_{em}\overline{m}_b}{8 \pi^2}
\bar{s}\sigma^{\mu\nu}(1+\gamma_5)b F_{\mu\nu}~,\hspace*{0.66cm}
Q_8=-\frac{g_{s}{\overline m}_b}{8 \pi^2}
\bar{s}\sigma^{\mu\nu}(1+\gamma_5) T^A b ~G^A_{\mu\nu}~,\\
Q_9&=\frac{\alpha_{em}}{2\pi} (\bar{s} b)_{V-A} (\bar{\ell}
\gamma^\mu \ell)~, \hspace*{1.96cm} Q_{10}=\frac{\alpha_{em}}{2\pi}
(\bar{s}b)_{V-A} (\bar{\ell} \gamma^\mu \gamma_5 \ell)~,
\end{aligned}
\end{equation}
where $T^A$ is the SU(3) color matrix, $\alpha_{em}=g^2_{em}/4\pi$ is
the fine-structure constant, and $\overline{m}_b(\mu)$ is the current mass of
the b quark in the $\overline{MS}$ scheme at the scale $\mu$.

Restricting ourselves to the kinematic region $1~\mbox{GeV}^2 <q^2 < 7
~\mbox{GeV}^2$, the light $K^\ast$ meson moves fast with a large
momentum of the order of $m_B/2$, which thus can be viewed approximately as
a collinear particle. For convenience, let us assume that the $K^\ast$
meson is moving in the direction of the light-like reference vector
$n$, then its momentum can be decomposed as $p^\mu= {\bar n}\cdot p n^\mu/2
+ p_\perp^\mu + n \cdot p {\bar n}^\mu/2$, where ${\bar n}^\mu$ is another
light-like reference vector satisfying $n \cdot {\bar n}=2$. In this
light-cone frame, the collinear momentum of $K^\ast$ is expressed as
\begin{equation}
p=(n \cdot p,{\bar n}\cdot p,p_\perp) \sim (\lambda^2, 1,
\lambda)m_b~,
\end{equation}
with $\lambda \sim \Lambda/m_b \ll 1$. In addition to this collinear mode, the
soft and hard-collinear modes, with momenta scaling as
$(\lambda,\lambda,\lambda)m_b$ and $(\lambda,1,\sqrt{\lambda})m_b$,
respectively, are also necessary to correctly reproduce the infrared
behavior of full QCD.

SCET introduces fields for every momentum mode and we will
encounter the following quark and gluon fields
\begin{eqnarray}
&&\xi_c \sim \lambda~,\hspace*{0.5cm} A_{c}^\mu \sim (\lambda^2, 1,
\lambda)~,\hspace*{0.5cm} \xi_{hc},~\xi_{\overline hc} \sim
\lambda^{1/2}~,\hspace*{0.5cm} A_{hc}^\mu \sim (\lambda, 1,
\lambda^{1/2})~, \nonumber \\
&&q_s \sim \lambda^{3/2}~,\hspace*{0.5cm} A_{s}^\mu \sim (\lambda,
\lambda, \lambda)~,\hspace*{0.5cm} h \sim \lambda^{3/2}~.
\end{eqnarray}
In the above, the symbol $A^\mu$ stands for the gluon field, $h$ represents a
heavy-quark field, the symbols $\xi$ and $q$ stand for the light quark fields, and
the subscripts c, s, hc stand for collinear, soft and hard-collinear modes, respectively.
Note that the momentum $q$ of the lepton pair is taken as a hard collinear
momentum, since in this paper we only consider the range
$1~\mbox{GeV}^2 <q^2 < 7 ~\mbox{GeV}^2$.
That is why an extra hard-collinear field $\xi_{\overline hc}$ in the
$\bar{n}$ direction is required later. As explained in detail
in Ref.~\cite{Neubert05}, to construct the gauge
invariant operators in SCET, it is more convenient to introduce the
building blocks, given below, which are obtained by multiplying the fields by
 the Wilson lines which run along the light-ray to infinity:
\begin{equation}
{\EuScript X}_{c}~,~~{\EuScript A}^\mu_{c}~,~~
{\EuScript X}_{hc}~,~~{\EuScript X}_{\overline hc}~,~~{\EuScript
A}^\mu_{hc}~,~~{\EuScript Q}_{s}~,~~{\EuScript
A}^\mu_{s}~,~~{\EuScript H}_{s}~,~~{\EuScript Q}_{\bar{s}}~,~~{\EuScript
H}_{\bar{s}}~.
\end{equation}
For example, the field ${\EuScript X}_{hc}$ is defined as
\begin{equation}
{\EuScript X}_{hc}(x)= W_{hc}^\dagger (x) \xi_{hc}(x);
~~{\rm with}~~W_{hc}(x) = P \exp \left(ig \int_{-\infty}^{0} ds \bar{n}\cdot A_{hc} (x + s \bar{n})\right),
\end{equation}
where $W_{hc}(x)$ is the hard collinear Wilson line.
The notations ${\EuScript Q}_{\bar{s}}$ and
$ {\EuScript H}_{\bar{s}} $ are used when the associated soft Wilson
lines are in the $\bar{n}$-direction.
For the definitions of the other fields and more technical details about SCET,
 we refer the reader to Ref.~\cite{Neubert05} and references therein.

Since SCET contains two kinds of collinear fields, {\it i.e.} hard-collinear
and collinear fields, normally an intermediate effective theory,
called $\mbox{SCET}_I$, is introduced which contains only soft and
hard-collinear fields. While the final effective theory, called
$\mbox{SCET}_{II}$, contains only soft and collinear fields. We will
then do a two-step matching from $QCD \to \mbox{SCET}_I \to
\mbox{SCET}_{II}$.

\subsection{QCD to $\mbox{SCET}_I$ matching}
In $\mbox{SCET}_I$, the $K^\ast$ meson is taken as a hard-collinear
particle and the relevant building blocks are ${\EuScript
X}_{hc}$, ${\EuScript X}_{\overline{hc}}$, ${\EuScript A}^\mu_{hc}$ and
$h$. The velocity of the B meson is
defined as $v=P_B/m_B$. The matching from QCD to $\mbox{SCET}_I$
at leading power may be expressed as
\begin{eqnarray}
H_{eff} \to & &-\frac{G_F}{\sqrt{2}} V_{ts}^\ast V_{tb} \left (
\sum_{i=1}^4 \int \! ds~\widetilde{C}_i^A (s) J_i^A (s) + \sum_{j=1}^4
\int \! ds \int \! dr~\widetilde{C}_j^B (s,r) J_j^B
(s,r) \right . \nonumber \\
& & \left . +\int \! ds \int \! dr \int \! dt
~\widetilde{C}^C(s,r,t) J^C(s,r,t)\right )~,
\end{eqnarray}
where $\widetilde{C}_i^{(A,B)}$ and $\widetilde{C}^C $ are Wilson coefficients in
the position space. The relevant $\mbox{SCET}_I$ operators
for the $B \to K^\ast \ell^+ \ell^-$ decay are constructed by using the
building blocks mentioned above~\cite{Neubert05}:
\begin{equation}
\begin{aligned}
J_1^A&=\bar{\EuScript X}_{hc}(s\bar{n})(1+\gamma_5)\gamma_\perp^\mu
h(0)~ \bar{\ell} \gamma_\mu \ell~,\hspace*{1cm} J_2^A=\bar{\EuScript
X}_{hc}(s\bar{n})(1+\gamma_5)
\frac{n^\mu}{n\cdot v} h(0)~ \bar{\ell} \gamma_\mu \ell~,  \\
J_3^A&=\bar{\EuScript X}_{hc}(s\bar{n})(1+\gamma_5)\gamma_\perp^\mu
h(0)~ \bar{\ell} \gamma_\mu \gamma_5 \ell~,\hspace*{0.6cm}
J_4^A=\bar{\EuScript X}_{hc}(s\bar{n})(1+\gamma_5)
\frac{n^\mu}{n\cdot v} h(0)~ \bar{\ell} \gamma_\mu \gamma_5 \ell~,   \\
J_1^B&=\bar{\EuScript X}_{hc}(s\bar{n})(1+\gamma_5)\gamma_\perp^\mu
\calAslash_{hc\perp}(r\bar{n})h(0)~ \bar{\ell} \gamma_\mu \ell~,  \\
J_2^B&=\bar{\EuScript
X}_{hc}(s\bar{n})(1+\gamma_5)\calAslash_{hc\perp}
(r\bar{n})\frac{n^\mu}{n\cdot v} h(0)~ \bar{\ell} \gamma_\mu \ell~, \\
J_3^B&=\bar{\EuScript X}_{hc}(s\bar{n})(1+\gamma_5)\gamma_\perp^\mu
\calAslash_{hc\perp}(r\bar{n})h(0)~ \bar{\ell} \gamma_\mu \gamma_5 \ell~,  \\
J_4^B&=\bar{\EuScript
X}_{hc}(s\bar{n})(1+\gamma_5)\calAslash_{hc\perp}
(r\bar{n})\frac{n^\mu}{n\cdot v} h(0)~ \bar{\ell} \gamma_\mu \gamma_5 \ell~,  \\
J^C&=\bar{\EuScript X}_{hc}(s\bar{n})(1+\gamma_5)\frac{\nbarslash}
{2} {\EuScript X}_{hc}(r\bar{n})~\bar{\EuScript
X}_{\overline hc}(an)(1+\gamma_5)
\frac{\nslash}{2} h(0)~,
\end{aligned}
\end{equation}
where the operators $J_i^A$ and $J_j^B$ represent the cases that the
lepton pair is emitted from the $b \to s$ transition currents, while
$J^C$ represents the diagrams in which the lepton pair is
emitted from the spectator quark of the B meson. Except
the lepton pair, the operators $J_i^{A,B}$ have the same Dirac
structures as those of the heavy-to-light transition currents in SCET,
which were first derived in Ref.~\cite{Bauer2} for $J_i^A$ and in
Refs.~\cite{Pirjol:2002km,Bauer:2002aj} for $J_j^B$ (see, also 
Refs.~\cite{Yang04,Neubert04}). In this paper we take the
operator basis of \cite{Neubert04,Neubert05} which makes $J_j^B$
multiplicatively renormalized, but we have neglected the operators which contain
the Dirac structure $\calAslash_{hc\perp}\gamma_\perp^\mu$ and which do not
contribute to the exclusive B meson decays. It is also clear that the
structure
$\bar{\ell} \gamma_\mu \gamma_5 \ell$ arises solely from $Q_{10}$ of
the weak effective Hamiltonian.

Since in practice the matching calculations are done
in the momentum space, it is more convenient to define the Wilson
coefficients in the momentum space by the following
Fourier-transformations:
\begin{equation}
\begin{aligned}
C_i^A(E)&=\int \! ds~e^{is\bar{n}\cdot P}
\widetilde{C}_i^A(s)~, \\
C_j^B(E,u)&=\int \! ds \int \! dr~e^{i(us+{\bar u}r){\bar n}\cdot P}
\widetilde{C}_j^B (s,r)~, \\
C^C(E,u)&=\int \! ds \int \! dr \int \! da~e^{i(us+{\bar
u}r){\bar n}\cdot P} e^{i an\cdot q} \widetilde{C}^C
(s,r,a)~,
\end{aligned}
\end{equation}
with $E\equiv n\cdot v {\bar n}\cdot P/2$ and ${\bar u}=1-u$. To get the
order $\alpha_s$ corrections to the decay amplitude, we need to
calculate the Wilson coefficients $C_i^A$ to one-loop level and
$C_j^{B}$ and $C^{C}$ to tree level. In the following we will use $\Delta_j
C_i^{(A,B,C)}$ to denote the matching results from the weak
effective operators $Q_j$ to the SCET currents $J_i^{A,B,C}$. With this,
the matching coefficients from $QCD \to \mbox{SCET}_I$ can be
written as
\begin{equation}
C_i^{(A,B,C)}= \sum_{j=1}^{10} \Delta_j
C_i^{(A,B,C)}(\mu_{QCD},\mu)~,
\end{equation}
where $\mu_{QCD}$ is the matching scale and $\mu$ is the
renormalization scale in $\mbox{SCET}_I$.

\begin{figure}[tb]
\begin{center}
\unitlength 1mm
\begin{picture}(140,81)
\put(0,0){\includegraphics{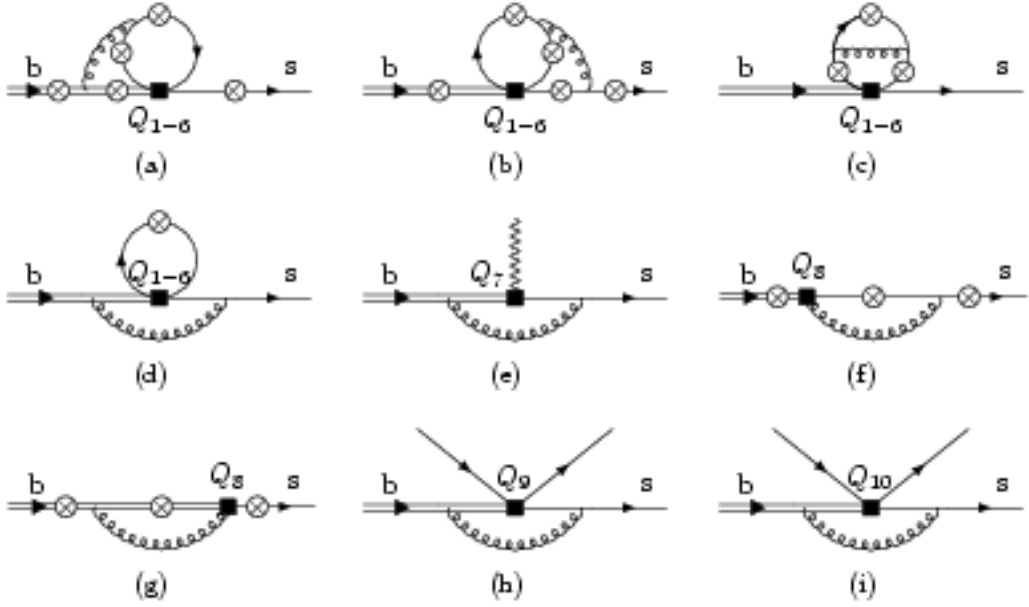}}
\end{picture}
\vspace*{-0.2cm}
\caption{${\cal O}(\alpha_s)$ contributions to the
matching of $Q_i$ to A-type SCET currents. The crossed circles denote
the possible locations from where the virtual photon is emitted and
then splits into a lepton pair.}
\end{center}
\end{figure}
Each operator of the weak effective Hamiltonian, namely $Q_{1-10}$,
will contribute to $C_i^A$ at order $\alpha_s$ level, as shown in
Fig. 1. But due to the small Wilson coefficients $C_{3-6}$, it
is numerically reasonable to neglect the contributions from
$Q_{3-6}$. For the operators $Q_{1,2}$ and $Q_8$, the results can be
easily derived from Eqs. (11) and (25) of Ref. \cite{Asatryan}:
\begin{equation}
\begin{aligned}
\Delta_{1,2} C_1^A(\mu_{\mathrm{\scriptscriptstyle
QCD}})&=-\frac{\alpha_{em}}{2\pi}\frac{\alpha_s(\mu_{\mathrm{\scriptscriptstyle
QCD}})}{4\pi}\left [
\frac{1}{\hat{s}}(2F_2^{(7)}+\hat{s}F_2^{(9)})\bar{C}_2 +
2(F_1^{(9)}+F_2^{(9)}/6)\bar{C}_1 \right ]~, \\
\Delta_{1,2} C_2^A(\mu_{\mathrm{\scriptscriptstyle
QCD}})&=-\frac{\alpha_{em}}{2\pi}\frac{\alpha_s(\mu_{\mathrm{\scriptscriptstyle
QCD}})}{4\pi}\left [ (2F_2^{(7)}+F_2^{(9)})\bar{C}_2 +
2(F_1^{(9)}+F_2^{(9)}/6)\bar{C}_1 \right ]~, \\
\Delta_8 C_1^A(\mu_{\mathrm{\scriptscriptstyle
QCD}})&=-\frac{\alpha_{em}}{2\pi}\frac{\alpha_s(\mu_{\mathrm{\scriptscriptstyle
QCD}})}{4\pi}\frac{\overline{m}_b(\mu_\mathrm{\scriptscriptstyle
QCD})}{m_b}
\left [ \frac{2}{\hat{s}}F_8^{(7)}+F_8^{(9)} \right ] C_8^{eff}~, \\
\Delta_8 C_2^A(\mu_{\mathrm{\scriptscriptstyle
QCD}})&=-\frac{\alpha_{em}}{2\pi}\frac{\alpha_s(\mu_{\mathrm{\scriptscriptstyle
QCD}})}{4\pi}\frac{\overline{m}_b(\mu_\mathrm{\scriptscriptstyle
QCD})}{m_b} \left [ 2 F_8^{(7)}+F_8^{(9)} \right ] C_8^{eff}~,
\end{aligned}
\end{equation}
where $\hat{s} \equiv q^2/m_b^2$ and $m_b$ is the pole mass of the b quark.
The current mass $\overline{m}_b$ is related to the pole mass at
next-to-leading order by
\begin{equation}
\overline{m}_b(\mu)=m_b \left [ 1+\frac{\alpha_s C_F}{4\pi} \left (
3\ln \frac{m_b^2}{\mu^2}-4 \right )  \right ]~,
\end{equation}
where $C_F=4/3$.
The functions $F_{1,2,8}^{(7,9)}$ are given in a mixed analytic and
numerical form in Ref. \cite{Asatryan}. Following the convention of Ref.
\cite{Beneke01}, we also use the "barred" coefficients
$\bar{C}_i$(i=1,...,6) here which are the linear combinations of the
Wilson coefficients $C_i$  of the weak effective Hamiltonian in
Eq.~(\ref{Heff}). The effective Wilson coefficient $C_8^{eff}$ is
defined as $C_8^{eff}=C_8+C_3-C_4/6+20C_5-10C_6$.

For the operators $Q_7$, $Q_9$ and $Q_{10}$, the matchings to the
A-type currents give
\begin{equation}
\begin{aligned}
\Delta_7 C_1^A&=\frac{\alpha_{em}}{2\pi}
\frac{\overline{m}_b(\mu_\mathrm{\scriptscriptstyle QCD})}{m_b}
\frac{2}{\hat{s}}\widetilde{C}_9 C_7^{eff}~, \hspace*{1cm} \Delta_7
C_2^A=\frac{\alpha_{em}}{2\pi}
\frac{\overline{m}_b(\mu_\mathrm{\scriptscriptstyle QCD})}{m_b}
2\widetilde{C}_{10} C_7^{eff}~, \\
\Delta_9 C_1^A&=\frac{\alpha_{em}}{2\pi}\widetilde{C}_3 C_9^{eff}~,
\hspace*{3cm}\Delta_9 C_2^A=\frac{\alpha_{em}}{2\pi}
\left ( \widetilde{C}_4+\frac{1-\hat{s}}{2}\widetilde{C}_5 \right )
C_9^{eff}~,\\
\Delta_{10}C_3^A&=\frac{\alpha_{em}}{2\pi}\widetilde{C}_3
C_{10}~,\hspace*{3.1cm} \Delta_{10}C_4^A=\frac{\alpha_{em}}{2\pi}
\left (\widetilde{C}_4+\frac{1-\hat{s}}{2}\widetilde{C}_5 \right )
C_{10}~.
\end{aligned}
\end{equation}
To avoid confusion with the Wilson coefficients in Eq.
(\ref{Heff}), we use the notations $\widetilde{C}_i$ for the
matching coefficients, instead of $C_i$ used originally in
Ref.~\cite{Bauer2}. The explicit expressions of $\widetilde{C}_i$ up
to one-loop order can be read from \cite{Bauer2,Yang04}. Note that
although the operator basis of the tensor current in \cite{Yang04}
looks slightly different from that of \cite{Bauer2}, they are
actually the same and it is easy to find the relations
$\widetilde{C}_9=C_T^{(A0)2}$ and $\widetilde{C}_{10}=C_T^{(A0)1}$.
The effective Wilson coefficients are defined as
$C_7^{eff}=C_7-C_3/3-4C_4/9-20C_5/3-80C_6/9$ and
$C_9^{eff}(q^2)=C_9+Y(q^2)$, where the function $Y(q^2)$ represents the
contributions of the fermion loops and the explicit formula can be
found in \cite{Beneke01}.

\begin{figure}[tb]
\begin{center}
\unitlength 1mm
\begin{picture}(140,54)
\put(0,0){\includegraphics{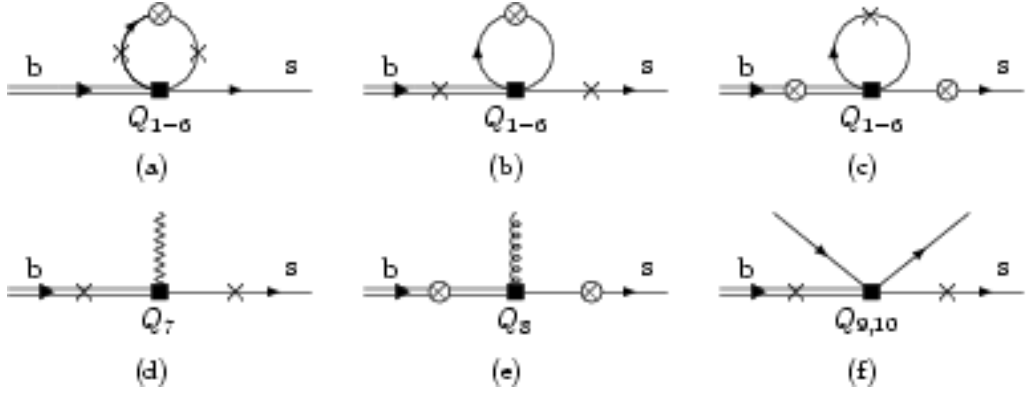}}
\end{picture}
\vspace*{-0.2cm} \caption{Tree-level matching of $Q_i$ onto B-type
SCET currents. The crossed circles denote the possible locations from
where the virtual photon is emitted, while the crosses mark the
possible places where a gluon line may be attached. }
\end{center}
\end{figure}
To get the decay amplitude of $B \to K^\ast \ell^+ \ell^-$ in order
$\alpha_s$, the tree-level matching of the effective weak
Hamiltonian (4) onto B-type SCET currents (11) is already enough, as
illustrated in Fig. 2. If we use the notation $\Delta_{16} C^B_i$ to
stand for the matchings of $Q_{1-6}$ onto B-type SCET currents
$J^B_i$, namely $\Delta_{16} C^B_i \equiv \sum_{j=1}^6 \Delta_j C^B_i$,
we get from Fig. 2a that
\begin{equation}
\begin{aligned}
\Delta_{16}C_1^B&=-\frac{\alpha_{em}}{2\pi}\frac{1}{m_b\hat{s}}
\left ( \frac{2}{3}
F_{16}^\perp(u,\hat{s},m_c^2/m_b^2)(\bar{C}_2+\bar{C}_4-\bar{C}_6)-\frac{1}{3}
F_{16}^\perp(u,\hat{s},0)\bar{C}_3 - \right . \\
&\hspace*{3cm}\left. \frac{1}{3}
F_{16}^\perp(u,\hat{s},1)(\bar{C}_3+\bar{C}_4-\bar{C}_6-4\bar{C}_5) \right
)~,
\\
\Delta_{16}C_2^B&=\frac{\alpha_{em}}{2\pi}\frac{2}{m_b} \left (
\frac{2}{3}
F_{16}^\parallel(u,\hat{s},m_c^2/m_b^2)(\bar{C}_2+\bar{C}_4-\bar{C}_6)-\frac{1}{3}
F_{16}^\parallel(u,\hat{s},0)\bar{C}_3- \right. \\
&\hspace*{3cm} \left. \frac{1}{3}
F_{16}^\parallel(u,\hat{s},1)(\bar{C}_3+\bar{C}_4-\bar{C}_6) \right )~,
\end{aligned}
\end{equation}
where $u$ is the momentum fraction carried by the strange quark in
the ${K}^\ast$ meson. The functions $F_{16}^{\perp,\parallel}$
are defined as
\begin{eqnarray}\label{F16perp}
F_{16}^\perp(u,\hat{s},\lambda)&=&1+\frac{2}{(1-\hat{s})(1-u)}\left (
\hat{s} \left ( \frac{\sqrt{-\hat{s}+4\lambda}}{\sqrt{\hat{s}~}}
\arctan
\frac{\sqrt{\hat{s}~}}{\sqrt{-\hat{s}+4\lambda}} - \hspace*{2.5cm}\right . \right . \nonumber \\
&&\left.
\frac{\sqrt{-1+u-\hat{s}u+4\lambda}}{\sqrt{1-(1-\hat{s})u}}\arctan
\frac{\sqrt{1-(1-\hat{s})u}}{\sqrt{-1+u-\hat{s}u+4\lambda}} \right
)+ \lambda~ \mbox{Li}_2 \left (
\frac{2\sqrt{\hat{s}~}}{\sqrt{\hat{s}~}-\sqrt{\hat{s}-4\lambda}}
\right ) \nonumber \\
&&+ \lambda~ \mbox{Li}_2 \left (
\frac{2\sqrt{\hat{s}~}}{\sqrt{\hat{s}~}+\sqrt{\hat{s}-4\lambda}}\right
) -\lambda~ \mbox{Li}_2 \left (
\frac{2\sqrt{1-(1-\hat{s})u}}{\sqrt{1-(1-\hat{s})u}+\sqrt{1-(1-\hat{s})u-4\lambda}}\right
) \nonumber \\
&& \left. -\lambda~ \mbox{Li}_2 \left (
\frac{2\sqrt{1-(1-\hat{s})u}}{\sqrt{1-(1-\hat{s})u}-\sqrt{1-(1-\hat{s})u-4\lambda}}\right
) \right )~,
\end{eqnarray}
\begin{eqnarray}\label{F16para}
F_{16}^\parallel(u,\hat{s},\lambda)&=&2\hat{s}+\frac{4\hat{s}}{(1-\hat{s})(1-u)}\left
( (1-u+u\hat{s}) \left (
\frac{\sqrt{-\hat{s}+4\lambda}}{\sqrt{\hat{s}~}} \arctan
\frac{\sqrt{\hat{s}~}}{\sqrt{-\hat{s}+4\lambda}} - \right . \right .
\nonumber \\
&&\left .\frac{\sqrt{-1+u-\hat{s}u+4\lambda}}{
\sqrt{1-(1-\hat{s})u}}\arctan
\frac{\sqrt{1-(1-\hat{s})u}}{\sqrt{-1+u-\hat{s}u+4\lambda}} \right
)+\lambda~ \mbox{Li}_2 \left (
\frac{2\sqrt{\hat{s}~}}{\sqrt{\hat{s}~}-\sqrt{\hat{s}-4\lambda}}
\right )
\nonumber \\
&& +\lambda~ \mbox{Li}_2 \left (
\frac{2\sqrt{\hat{s}~}}{\sqrt{\hat{s}~}+\sqrt{\hat{s}-4\lambda}}\right
) - \lambda~ \mbox{Li}_2 \left (
\frac{2\sqrt{1-(1-\hat{s})u}}{\sqrt{1-(1-\hat{s})u}+\sqrt{1-(1-\hat{s})u-4\lambda}}\right
)\nonumber \\
&& - \left. \lambda~ \mbox{Li}_2 \left (
\frac{2\sqrt{1-(1-\hat{s})u}}{\sqrt{1-(1-\hat{s})u}-\sqrt{1-(1-\hat{s})u-4\lambda}}\right
) \right )~.
\end{eqnarray}
As a check, it is not difficult to find the following relations
\[
F_{16}^\perp(u,\hat{s},\frac{m_q^2}{m_b^2})=t_\perp(u,m_q)\times \frac{(1-u)E}{2M_B}~,
\hspace*{1cm} F_{16}^\parallel(u,\hat{s},\frac{m_q^2}{m_b^2})=t_\parallel(u,m_q)\times
\frac{\hat{s}(1-u)E}{M_B}~,
\]
where the functions $t_{\perp,\parallel}(u,m_q)$ are defined in
Eqs.~(27)-(28) in the paper by Beneke
{\it et al.}~\cite{Beneke01}. We also note that the functions
$F_{16}^\perp(u,\hat{s},\lambda)$ and $F_{16}^\parallel(u,\hat{s},\lambda)$
are finite as $\bar{u}=1-u \to 0$, as opposed to the functions
$t_{\perp,\parallel}(u,m_q)$, which are singular as $\bar{u} \to 0$.

Fig.~2d and the operator $Q_9$ of Fig.~2f, combined with Fig.~2b,
will contribute to the matching coefficients
$\Delta_{7,9}C_{1,2}^B$, while the operator $Q_{10}$ of Fig.~2f will
contribute to $\Delta_{10} C_{3,4}^B$:
\begin{equation}
\begin{aligned}
\Delta_7 C_1^B&=-\frac{\alpha_{em}}{2\pi}\frac{\overline{m}_b}{m_b^2
\hat{s}} 2 C_7^{eff}~, \hspace*{0.5cm}
\Delta_7C_2^B=\frac{\alpha_{em}}{2\pi}\frac{\overline{m}_b}{m_b^2
(1-\hat{s})} 2 C_7^{eff}~, \\
\Delta_9 C_1^B&=0~, \hspace*{3.2cm}
\Delta_9C_2^B=-\frac{\alpha_{em}}{2\pi}\frac{1-2\hat{s}}{m_b(1-\hat{s})}
C_9^{eff}~, \\
\Delta_{10} C_3^B&=0~, \hspace*{3.05cm}
\Delta_{10}C_4^B=-\frac{\alpha_{em}}{2\pi}\frac{1-2\hat{s}}{m_b(1-\hat{s})}
C_{10}~.
\end{aligned}
\end{equation}
Finally, Fig.~2e and Fig.~2c contribute to the matching
coefficients
\begin{equation}
\Delta_8 C_1^B=-\frac{\alpha_{em}}{2\pi}\frac{\overline{m}_b}{m_b^2}
\frac{2(1-u)(1-\hat{s})}{3\hat{s}(u+\hat{s}-u \hat{s})}C_8^{eff}~,
\hspace*{1.5cm} \Delta_8 C_2^B=0~.
\end{equation}
\begin{figure}[tb]
\begin{center}
\unitlength 1mm
\begin{picture}(140,73)
\put(0,0){\includegraphics{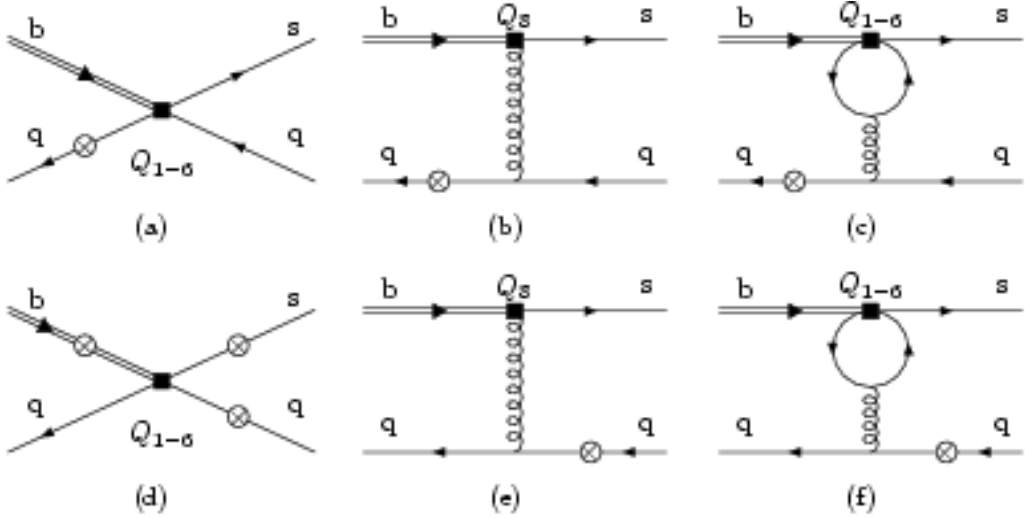}}
\end{picture}
\vspace*{-0.2cm} \caption{The diagrams where
the virtual photon, as
denoted by the crossed circle, is emitted from the spectator quark.}
\end{center}
\end{figure}
We shall now consider the diagrams where
the virtual (off-shell) photon is emitted from the spectator quark, as shown in
Fig.~3. Due to the off-shellness of the quark propagator, it is easy
to check that Fig.~(3d-3f) are of  order $1/m_b$ suppressed compared
with Fig.~(3a-3c) where the photon is emitted from the spectator
quark in the B meson. Therefore at leading power in $1/m_b$, only the
first three diagrams in Fig.~3 are relevant for our analysis. As we
shall see in the following, all of these three diagrams contribute
to the Wilson coefficients of the C-type SCET current.

The annihilation diagram, shown in Fig.~3a, contributes to the matching
coefficient $C^C$ at order $\alpha_s^0$, for which the calculation is
trivial,
\begin{equation}\label{anni1}
\Delta_{16}^{(0)} C^C =
\frac{2}{3}\left ( -\frac{V_{us}^\ast V_{ub}}{V_{ts}^\ast
V_{tb}}(\bar{C}_1+3\bar{C}_2)\delta_{qu}+(\bar{C}_3+3\bar{C}_4)\right
)~.
\end{equation}
Here $q$ is the flavor of the spectator quark in the B meson and the
superscript $(0)$ denotes the matching at order
$\alpha_s^0$. At order $\alpha_s$, the diagrams shown in Figs.~(3b-3c) also contribute to
the matching onto the C-type SCET current with the coefficients
\begin{equation}\label{anni2}
\begin{aligned}
\Delta_8
C^C&=\frac{C_F}{N_c}\frac{\alpha_s}{4\pi}
\frac{-4C_8^{eff}}{1-u+u\hat{s}}~, \\
\Delta_{16}^{(1)} C^C &=
2\frac{C_F}{N_c}\frac{\alpha_s}{4\pi} \left \{ \frac{}{}
(\bar{C}_2+\bar{C}_4+\bar{C}_6)G(u,\hat{s},m_c^2/m_b^2)+
(\bar{C}_3+3\bar{C}_4+3\bar{C}_6)G(u,\hat{s},0) \right .\\
 &\hspace*{2.5cm} \left. +
(\bar{C}_3+\bar{C}_4+\bar{C}_6)G(u,\hat{s},1)+
 \frac{4}{9}(\bar{C}_3-\bar{C}_5-15\bar{C}_6)
\right \}~,
\end{aligned}
\end{equation}
where the function $G(u,\hat{s},\lambda)$ is defined as
\begin{equation}
G(u,\hat{s},\lambda)=\frac{2}{3}+\frac{2}{3}\ln \frac{m_b^2}{\mu^2}+
4\int_0^1 \! dx~x(1-x)\ln [\lambda-x(1-x)(1-u+u\hat{s}) ]~.
\end{equation}
%
\subsection{$\mbox{SCET}_I \to \mbox{SCET}_{II}$ matching}
As shown in Refs.~\cite{Bauer:2002aj,Neubert04B}, which analyzed the form factors
in the framework of SCET, one may simply define the matrix
elements of the A-type $\mbox{SCET}_I$ currents as non-perturbative
input since the non-factorizable parts of the form factors are all
contained in such matrix elements. Therefore the explicit matching
of $J_i^A$ to $\mbox{SCET}_{II}$ operators is not necessary here.

For B-type $\mbox{SCET}_I$ operators, they are matched onto the
following $\mbox{SCET}_{II}$ operators
\begin{equation}
\begin{aligned}
O_1^B&=\bar{\EuScript X}_{c}(s\bar{n})(1+\gamma_5)\gamma_\perp^\mu
\frac{\nbarslash}{2}{\EuScript X}_{c}(0)~\bar{\EuScript Q}_{s}(t
n)(1-\gamma_5)\frac{\nslash}{2}{\EuScript H}_{s}(0)~ \bar{\ell}
\gamma_\mu \ell~, \\
O_2^B&=\bar{\EuScript
X}_{c}(s\bar{n})(1+\gamma_5)\frac{n^\mu}{n\cdot
v}\frac{\nbarslash}{2}{\EuScript X}_{c}(0)~\bar{\EuScript Q}_{s}(t
n)(1+\gamma_5)\frac{\nslash}{2}{\EuScript H}_{s}(0)~ \bar{\ell}
\gamma_\mu \ell~, \\
O_3^B&=\bar{\EuScript X}_{c}(s\bar{n})(1+\gamma_5)\gamma_\perp^\mu
\frac{\nbarslash}{2}{\EuScript X}_{c}(0)~\bar{\EuScript Q}_{s}(t
n)(1-\gamma_5)\frac{\nslash}{2}{\EuScript H}_{s}(0)~ \bar{\ell}
\gamma_\mu \gamma_5 \ell~, \\
O_4^B&=\bar{\EuScript
X}_{c}(s\bar{n})(1+\gamma_5)\frac{n^\mu}{n\cdot
v}\frac{\nbarslash}{2}{\EuScript X}_{c}(0)~\bar{\EuScript Q}_{s}(t
n)(1+\gamma_5)\frac{\nslash}{2}{\EuScript H}_{s}(0)~ \bar{\ell}
\gamma_\mu \gamma_5 \ell~,
\end{aligned}
\end{equation}
where we only include the color-singlet operators that have non-zero
matrix elements for the $B \to K^\ast \ell^+ \ell^-$ decay. Again,
it is in practice more convenient to do the matching calculations in
the momentum space, and the Wilson coefficients $D_i^B(\omega,u)$ can be
defined by Fourier transforming the corresponding ones
 $\tilde{D}_i^B(s,t)$ introduced in
the position space, just like the case in $\mbox{SCET}_I$,
\begin{equation}
D_i^B(\omega,u)=\int \! ds \int \! dt~ e^{-i\omega n\cdot v t}
e^{ius\bar{n}\cdot P} \tilde{D}_i^B (s,t).
\end{equation}
Following the notations of \cite{Neubert05}, the Wilson coefficients
$D_i^B$ can be expressed as
\begin{equation}
D_i^B(\omega, u, \hat{s}, \mu)=\frac{1}{\omega}\int_0^1 \! dv~{\cal J}_i \left
( u,v,\ln\frac{m_b \omega (1-\hat{s})}{\mu^2},\mu \right )
C_i^B(v,\mu)~,
\end{equation}
where the jet functions ${\cal J}_i$ arise from the $\mbox{SCET}_I
\to \mbox{SCET}_{II}$ matching and it is clear that ${\cal
J}_1={\cal J}_3\equiv {\cal J}_\perp$ and ${\cal J}_2={\cal
J}_4\equiv {\cal J}_\parallel$. At tree level, using the
Fierz transformation in the operator basis,
\begin{equation}
\begin{aligned}
\bar{\EuScript X}_c N {\EuScript H}_{s}~\bar{\EuScript Q}_{s} M
{\EuScript X}_c &=-\frac{1}{4}\bar{\EuScript
X}_c(1+\gamma_5)\frac{\nbarslash}{2}{\EuScript X}_{c}~
\bar{\EuScript Q}_{s}M(1-\gamma_5)\frac{\nslash}{2}N{\EuScript
H}_{s}-\frac{1}{4}\bar{\EuScript
X}_c(1-\gamma_5)\frac{\nbarslash}{2}{\EuScript X}_{c} \\
& \times \bar{\EuScript Q}_{s}M(1+\gamma_5)\frac{\nslash}{2}N{\EuScript
H}_{s}-\frac{1}{8}\bar{\EuScript
X}_c(1+\gamma_5)\frac{\nbarslash}{2}\gamma_{\perp \alpha}{\EuScript
X}_{c}~ \bar{\EuScript Q}_{s}M(1+\gamma_5)\gamma_\perp^\alpha
\frac{\nslash}{2}N{\EuScript H}_{s}~,
\end{aligned}
\end{equation}
one obtains
\begin{equation}
{\cal J}_\perp(u,v)={\cal J}_\parallel(u,v)=-\frac{4\pi C_F
\alpha_s}{N_c}\frac{1}{m_b(1-u)(1-\hat{s})}\delta (u-v)~.
\end{equation}
Finally, the C-type $\mbox{SCET}_I$ current is matched
onto the $\mbox{SCET}_{II}$ operator
\begin{equation}
O^C=\bar{\EuScript X}_{c}(s\bar{n})(1+\gamma_5)\frac{\nbarslash}{2}
{\EuScript X}_{c}(0)~\bar{\EuScript Q}_{\bar s}(t \bar
{n})(1+\gamma_5)\frac{\nslash}{2} {\EuScript
H}_{\bar s}(0)~ \frac{\bar{n}^\mu}{\bar{n}\cdot v}\bar{\ell} \gamma_\mu 
\ell~.
\end{equation}
We may similarly define
\begin{equation}
D^C(\omega,u)=\int \! ds \int \! dt~ e^{-i\omega \bar{n}\cdot v t}
e^{ius\bar{n}\cdot P} \tilde{D}^C (s,t).
\end{equation}
with
\begin{equation}
D^C(\omega, u, \hat{s}, \mu)=\frac{-e
e_q \hat{s}}{(\omega-q^2/m_b-i\epsilon)}{\cal J}^C
\left
(\ln\frac{m_b \omega (1-\hat{s})}{\mu^2},\mu \right )
C^C(E,u,\mu)~,
\end{equation}
where $e_q$ is the electric charge of the spectator quark in the B
meson. At tree level the corresponding jet function is
trivial, ${\cal J}^C=1$. For later convenience, we will define
$D^C \equiv \widehat{D}^C/(\omega-q^2/m_b-i\epsilon)$.
\subsection{Matrix elements of SCET operators}
The last step before we can finally get the decay amplitude for the $B
\to K^\ast \ell^+ \ell^-$ decay is to take the matrix elements of
the relevant SCET operators. For the A-type SCET currents (11), one may
simply define~\cite{Neubert05}
\begin{equation}\label{SCETff}
\langle M(p)\vert \bar{\EuScript X}_{hc}\Gamma h \vert
B(v)\rangle=-2E\zeta_M(E)tr[\overline{\cal M}_M(n)\Gamma {\cal
M}_B(v)]~,
\end{equation}
where the projection operators are
\begin{equation}
{\cal M}_B(v)=-\frac{1+\rlap{\hspace{0.02cm}/}v}{2} \gamma_5~,
\hspace*{1cm} \overline{\cal M}_{K^\ast_\perp}(n)=\polslash^*_\perp
\frac{\nbarslash \nslash}{4}~,\hspace*{1cm} \overline{\cal
M}_{K^\ast_\parallel}(n)=-\frac{\nbarslash \nslash}{4}~,
\end{equation}
with $\varepsilon^\mu_\perp$ being the polarization vector of the
 $K^\ast_\perp$ meson. It is then straightforward to get the matrix elements
 of the $\mbox{SCET}_I$ currents $J_i^A$ as
\begin{equation}
\begin{aligned}
\langle K^\ast \ell^+ \ell^- \vert J_1^A \vert B \rangle &=-2E
\zeta_\perp(g^{\mu\nu}_\perp - i\epsilon^{\mu\nu}_\perp )
\varepsilon^{*}_{\perp\nu} \bar{\ell}\gamma_\mu \ell~,
\hspace*{0.9cm} \langle K^\ast \ell^+ \ell^- \vert J_2^A \vert B
\rangle =-2E \zeta_\parallel
\frac{n^\mu}{n\cdot v} \bar{\ell}\gamma_\mu \ell~, \\
\langle K^\ast \ell^+ \ell^- \vert J_3^A \vert B \rangle &=-2E
\zeta_\perp(g^{\mu\nu}_\perp - i\epsilon^{\mu\nu}_\perp )
\varepsilon^{*}_{\perp\nu} \bar{\ell}\gamma_\mu \gamma_5 \ell~,
\hspace*{0.5cm} \langle K^\ast \ell^+ \ell^- \vert J_4^A \vert B
\rangle =-2E \zeta_\parallel \frac{n^\mu}{n\cdot v}
\bar{\ell}\gamma_\mu \gamma_5 \ell~,
\end{aligned}
\end{equation}
where $g^{\mu\nu}_\perp \equiv g^{\mu\nu}-(n^\mu {\bar n}^\nu+{\bar
n}^\mu n^\nu)/2$ and $\epsilon^{\mu\nu}_\perp\equiv
\epsilon^{\mu\nu\rho\sigma}v_\rho n_\sigma/(n\cdot v)$. Note that in the
above equations, we use the convention $\epsilon^{0123}=+1$, as adopted in
the book by Peskin and Schroeder \cite{Peskin}.

For the B-type $\mbox{SCET}_{II}$ operators (25), although naively the soft
and collinear degrees of freedom seem to be decoupled, the factorization may
be invalidated unless no endpoint divergences appear in the
convolution integrals \cite{Beneke04,Neubert04B}. The relevant meson
LCDAs are defined as \cite{BBNS,Neubert97}
\begin{equation}
\begin{aligned}
\langle 0 \vert \bar{\EuScript Q}_{s}(tn)\Gamma {\EuScript H}_{
s}(0) \vert B(v)\rangle=&\frac{iF(\mu)}{2}\sqrt{m_B}\int_0^\infty \!
d\omega~e^{-i\omega n\cdot v t} \\
&\hspace*{-0.5cm}tr \left [ \left (
\phi^B_+(\omega,\mu)-\frac{\nslash}{2n\cdot v}
(\phi^B_-(\omega,\mu)-\phi^B_+(\omega,\mu))\right ) \Gamma
{\cal M}_B(v) \right ]~,\\
\langle K^\ast(p) \vert \bar{\EuScript X}_c(s\bar{n})\Gamma
\frac{\nbarslash}{2}{\EuScript X}_c(0) \vert 0 \rangle =&
\frac{if_{K^\ast}(\mu)}{4}\bar{n}\cdot p~ tr[\overline{\cal
M}_{K^\ast}\Gamma]\int_0^1 \! du~e^{ius\bar{n}\cdot p}
\phi_{K^\ast}(u,\mu)~,
\end{aligned}
\end{equation}
where two different $K^*$-distribution amplitudes
($\phi_{K^*}^\parallel(u,\mu)  $
for $\Gamma=1$ and $\phi_{K^*}^\perp(u,\mu)  $
for $\Gamma=\gamma_\perp$) with their corresponding decay constants
$f_{K^*}^\parallel$ and 
$f_{K^*}^\perp(\mu)$, respectively, are involved;  
 $F(\mu)$ is related to the B meson decay constant $f_B$ up to higher
orders in $1/m_b$ by \cite{Neubert94}
\begin{equation}
 f_B \sqrt{m_B}=F(\mu) \left ( 1+\frac{C_F \alpha_s(\mu)}{4\pi}
                            \left ( 3\ln\frac{m_b}{\mu}-2 \right )
\right )~.
\end{equation}
With the above LCDAs, the matrix elements of the operators $O_i^B$
can be written as
\begin{equation}
\begin{aligned}
\langle K^\ast \ell^+ \ell^- \vert C_1^B O_1^B \vert B \rangle
&=-\frac{F(\mu)m_B^{3/2}}{4}(1-\hat{s})(g^{\mu\nu}_\perp -
i\epsilon^{\mu\nu}_\perp )
\varepsilon^{*}_{\perp\nu}\bar{\ell}\gamma_\mu \ell~\int_0^\infty
\frac{d\omega}{\omega}\phi_+^B(\omega,\mu)\\ & \times \int_0^1 \! du~
f_{K^\ast_\perp}(\mu)\phi_{K^\ast_\perp}(u,\mu) \int_0^1 \! dv {\cal
J}_\perp(u,v,\ln \frac{m_b \omega
(1-\hat{s})}{\mu^2},\mu)C_1^B(v,\mu) \\
&\equiv-\frac{F(\mu)m_B^{3/2}}{4}(1-\hat{s})(g^{\mu\nu}_\perp -
i\epsilon^{\mu\nu}_\perp )
\varepsilon^{*}_{\perp\nu}\bar{\ell}\gamma_\mu \ell~ \phi_+^B
\otimes f_{K^\ast_\perp}\phi_{K^\ast_\perp} \otimes {\cal J}_\perp
\otimes C_1^B~,\\
\langle K^\ast \ell^+ \ell^- \vert C_2^B O_2^B \vert B
\rangle&=-\frac{F(\mu)m_B^{3/2}}{4}(1-\hat{s}) \frac{n^\mu}{n\cdot
v}\bar{\ell}\gamma_\mu \ell~ \phi_+^B \otimes
f_{K^\ast_\parallel}\phi_{K^\ast_\parallel} \otimes {\cal
J}_\parallel \otimes C_2^B~,
\end{aligned}
\end{equation}
while for the matrix element of $C_3^B O_3^B$($C_4^B O_4^B$), it can
be obtained by simply replacing the lepton current
$\bar{\ell}\gamma_\mu \ell$ on the right hand side of the above
equations by $\bar{\ell}\gamma_\mu \gamma_5 \ell$ and also
replacing $C_1^B
\to C_3^B$ ($C_2^B \to C_4^B$).

The matrix element of $O^C$ is obtained likewise, with the result
\begin{equation}
\langle K^\ast \ell^+ \ell^- \vert D^C O^C \vert B
\rangle=-\frac{F(\mu)m_B^{3/2}}{4}(1-\hat{s})\frac{\bar{n}^\mu}
{\bar{n}\cdot
v}\bar{\ell} \gamma_\mu \ell~\frac{\omega \phi_-^B}{\omega-q^2/m_b-i\epsilon}
\otimes f_{K^\ast_\parallel} \phi_{K^\ast_\parallel} \otimes
\widehat{D}^C~.
\end{equation}
Since $\phi_-^B(\omega)$ does not vanish as $\omega$ approaches
zero, the integral $\int \!
d\omega~\phi_-^B(\omega)/(\omega-q^2/m_b)$ would be divergent if
$q^2\to 0$. This endpoint singularity will violate the
$\mbox{SCET}_{II}$ factorization, that is why we should restrict our
attention to the kinematic region where the invariant mass of the
lepton pair is not too small, say $q^2\geq 1~\mbox{GeV}^2$.
\subsection{Resummation of logarithms in SCET}
In the above analysis a two-step matching procedure $\mbox{QCD} \to
\mbox{SCET}_I \to \mbox{SCET}_{II}$ has been implemented. This
introduces two matching scales, $\mu_h \sim m_b$ at which QCD is
matched onto SCET$_I$ and $\mu_l \sim \sqrt{m_b \Lambda}$ at which
SCET$_I$ is matched onto SCET$_{II}$. Thus, with the SCET$_I$
matching coefficients at scale $\mu_h$, one may use the
renormalization-group equations (RGE) of SCET$_I$ to evolve them
down to scale $\mu_l$ and then match onto SCET$_{II}$. The large
logarithms due to different scales are resummed during this procedure.
Note that the meson LCDAs may be given at another scale $\mu_L$,
and, in principle, one should also use the RGE of SCET$_{II}$ to run the
corresponding matching coefficients from $\mu_l$ down to $\mu_L$.
But since in B decays the scale $\mu_l \simeq 1.5 \mbox{GeV}$ is
already quite low, we may just take the meson LCDAs at the scale $\mu_l$
in this paper for simplicity and thereby avoid the running of the
SCET$_{II}$ matching coefficients.

Furthermore, one should note that for the A-type SCET currents, only the
scale $\mu_h$ is involved since it is not necessary to do the second step
matching of SCET$_I \to$ SCET$_{II}$. Similarly, we
may choose the nonperturbative form factors $\zeta_{\perp,\parallel}$
at the scale $\mu_h$ and avoid the RGE running of the A-type SCET$_I$
matching coefficients. For the B-type currents, the RGE of SCET$_I$ can
be obtained by calculating the anomalous dimensions of the relevant
SCET operators, which has been done in \cite{Neubert04}, where the
matching coefficients at any scale $\mu$ can be obtained by an evolution from
the matching scale $\mu_h$ as follows
\begin{equation}\label{evolution}
\begin{aligned}
C^B_j (E,u,\mu_h, \mu)&=\left ( \frac{2E}{\mu_h} \right )^{a(\mu_h,\mu)}
e^{S(\mu_h,\mu)} \int_0^1 \! dv~U_\Gamma (u,v,\mu_h,\mu)C_j^B
(E,v,\mu_h)\\
&\equiv \left ( \frac{2E}{\mu_h} \right )^{a(\mu_h,\mu)}
e^{S(\mu_h,\mu)}~ \widetilde{U}^j_\Gamma (E,u,\mu_h,\mu)~,
\end{aligned}
\end{equation}
with the subscript $\Gamma=\perp,\parallel$ and the functions
$a(\mu_h,\mu)$ and $S(\mu_h,\mu)$ are given in Eq.~(66) of Ref.~\cite{Neubert04}. Note
that in the above equation one should use the subscript $\Gamma=\perp$
for $j=1,3$, while $\Gamma=\parallel$ for $j=2,4$. The evolution kernel
$\widetilde{U}^j_\Gamma (E,u,\mu_h,\mu)$ obeys
\begin{equation}\label{kernel}
\frac{d\widetilde{U}^j_\Gamma (E,u,\mu_h,\mu)}{d\ln\mu}=\int_0^1 \!
dy~ y V_\Gamma (y,u)\widetilde{U}^j_\Gamma (E,y,\mu_h,\mu) +
\omega(u)\widetilde{U}^j_\Gamma (E,u,\mu_h,\mu)~,
\end{equation}
with the initial condition
$\widetilde{U}^j_\Gamma (E,u,\mu_h,\mu_h)=C_j^B(E,u,\mu_h)$.
Again, the functions $V_\Gamma (y,u)$ and
$\omega(u)$ are defined in \cite{Neubert04}. In the next section on
phenomenological application, we will solve the above
integro-differential equation numerically.

Finally, for the C-type SCET current $J^C$, its anomalous dimension
just equals the sum of the anomalous dimensions of the $K^\ast$
meson LCDA $\phi_{K^*}$ and the B meson LCDA $\phi_-^B$. However, as the
evolution equation of $\phi_-^B$ is still unknown, we will not
resum the  perturbative logarithms for the $J^C$ current in this
paper. Numerically the contribution from the $J^C$ current to the decay
amplitude is small. Furthermore, as we will see later, the $J^C$ current is
completely irrelevant for the forward-backward asymmetry of the
charged leptons. Therefore, this treatment has only minor
impact on our phenomenological discussion.
\section{Numerical analysis of $B \to K^\ast \ell^+ \ell^-$}
We are now in the position to write the decay amplitude of $B \to
K^\ast \ell^+ \ell^-$, using the similar notations adopted in
\cite{Beneke01},
\begin{equation}\label{decayamplitude}
\begin{aligned}
\frac{d^2 \Gamma}{d q^2 d \cos \theta} &= \frac{G_F^2 \vert
V_{ts}^\ast V_{tb} \vert^2}{128\pi^3} \left
(\frac{\alpha_{em}}{4\pi} \right )^2 m_B^3 \lambda_{K^\ast}
 (1-\frac{q^2}{m_B^2})^2 \times \\
& \left \{ 2\zeta_\perp^2 (1+\cos^2 \theta)\frac{q^2}{m_B^2} (\vert
{\cal C}_{9}^\perp \vert^2 + ({\cal C}_{10}^\perp)^2 )
 \right. \\
& \left . - 8\zeta_\perp^2 \cos \theta \frac{q^2}{m_B^2} Re({\cal
C}_{9}^\perp){\cal C}_{10}^\perp +\zeta_\parallel^2(1-\cos^2
\theta)(\vert {\cal C}_{9}^\parallel \vert^2 + ({\cal
C}_{10}^\parallel)^2 ) \right \}~,
\end{aligned}
\end{equation}
with $m_B \lambda_{K^\ast}/2$ being the 3-momentum of the $K^\ast$ meson in the
rest frame of the B meson,
\begin{equation}
 \lambda_{K^\ast} =\left [ \left ( 1-\frac{q^2}{m_B^2}
\right )^2 - 2 \frac{m^2_{K^\ast}}{m_B^2} \left (
1+\frac{q^2}{m_B^2} \right ) + \frac{m^4_{K^\ast}}{m_B^4} \right
]^{1/2}~.
\end{equation}
The angle $\theta$ denotes the angle between the momenta of the
positively charged lepton and  the B meson in the rest frame
of the lepton pair. Note that in the above equations the leptons are
taken in the massless limit and the $K^\ast$ meson mass is kept nonzero
only for $\lambda_{K^\ast}$, which arises from the phase space.
The "effective" Wilson coefficients ${\cal C}_9^{\perp,\parallel}$ and ${\cal
C}_{10}^{\perp,\parallel}$ are given by

\begin{equation}
\begin{aligned}
{\cal C}_9^\perp &= \frac{2\pi}{\alpha_{em}} \left ( C_1^A +
\frac{m_B}{4}\frac{f_B \phi_+^B \otimes
f^\perp_{K^\ast}\phi^\perp_{K^\ast} \otimes {\cal J}_\perp \otimes
C_1^B}{\zeta_\perp} \right )~, \\
{\cal C}_9^\parallel &= \frac{2\pi}{\alpha_{em}} \left (C_2^A +
\frac{m_B}{4}\frac{f_B \phi_+^B \otimes
f^\parallel_{K^\ast}\phi^\parallel_{K^\ast} \otimes {\cal
J}_\parallel \otimes C_2^B}{\zeta_\parallel}  \right .\\
& \left . \hspace*{2.cm} - \frac{q^2}{4m_B}
\frac{f_B\omega \phi_-^B/(\omega-q^2/m_b-i\epsilon)
\otimes f_{K^\ast}^\parallel \phi_{K^\ast}^\parallel \otimes
\widehat{D}^C}{\zeta_\parallel}\right )~,\\
{\cal C}_{10}^\perp &= \frac{2\pi}{\alpha_{em}}C_3^A~,\\
{\cal C}_{10}^\parallel &= \frac{2\pi}{\alpha_{em}} \left (C_4^A +
\frac{m_B}{4}\frac{f_B \phi_+^B \otimes
f^\parallel_{K^\ast}\phi^\parallel_{K^\ast} \otimes {\cal
J}_\parallel \otimes C_4^B}{\zeta_\parallel} \right )~,
\end{aligned}
\end{equation}
where $C_i^{\rm A,B}$ and $D^{\rm C} $ are defined in Eqs.~(13)
and (32), respectively.
The above expressions are valid at leading power in $1/m_b$ and to
all orders in $\alpha_s$. But in this paper we only calculate explicitly
the "effective Wilson coefficients" at one-loop order. At this order our
results are quite similar to those of \cite{Beneke01} using the
large-energy limit of QCD.
The main phenomenological improvement is that for
the hard scattering part, the matching coefficients $C_i^B$ are evolved
from the scale $\mu_h \sim {\cal O}(m_b)$ down to $\mu_l \sim
\sqrt{m_b \Lambda_h}$, during which the perturbative logarithms
are summed. Here, $\Lambda_h$
represents a typical hadronic scale. Note also that the
definitions of the soft form factors $\zeta_{\perp,\parallel}$ in SCET are
different from those of Ref \cite{Beneke01},
therefore the explicit expressions for $C^A_i$ are also different from the
coefficients $C_a^{0,1}$ appearing in
\cite{Beneke01} which are related to the form factor corrections.

In terms  of the helicity amplitudes for the decay $B \to K^* (\to K + \pi)
\ell^+\ell^- $, the double differential distribution $d^2{\cal B}/d\cos
\theta_+ ds$ is given in Eq.(44) of Ref. \cite{AliSafir}. This requires the
helicity amplitudes, $\vert H_0(s)\vert^2= \vert H^L_0(s)\vert^2 + \vert
H^R_0(s)\vert^2$, $ \vert H_-^{L,R}(s)\vert^2$ and $ \vert
H_+^{L,R}(s)\vert^2$.
While the amplitudes $H_+^{L,R}(s)$ are both power suppressed in $1/m_b$ and
numerically small, the expressions for the others in SCET are given below:
\begin{equation}
\begin{aligned}
\vert H_0 \vert^2&=\frac{m_B^2}{2}(1-\frac{q^2}{m_B^2})^2
(\vert {\cal C}_{9}^\parallel \vert^2 + ({\cal
C}_{10}^\parallel)^2 ) \zeta_\parallel^2~,\\
\vert H_-^{L,R} \vert^2&= q^2 (1-\frac{q^2}{m_B^2})^2
\vert {\cal C}_{9}^\perp \pm {\cal
C}_{10}^\perp \vert^2  \zeta_\perp^2~.
\end{aligned}
\end{equation}
Note that the dependence on the soft form factors factorizes in
$\zeta_\parallel^2$ and $ \zeta_\perp^2$ for the helicity components
$ \vert H_0 \vert^2 $ and $ \vert H_-^{L,R} \vert^2 $, respectively.
Since a similar analysis in terms of the helicity amplitudes of the charged
current decay $B \to \rho (\to \pi \pi) \ell^+\nu_\ell$ can be performed, the
ratios $R_{0}(s)$ and $R_{-}(s)$ of the two differential distributions
(in $B \to K^* (\to K \pi) \ell^+ \ell^-$ and $B \to \rho (\to \pi \pi) \ell^+\nu_\ell$)
have lot less hadronic uncertainties, as these ratios (see Eq. (76)
 in Ref.~\cite{AliSafir} for their definition) involve estimates of the
 SU(3)-breaking in the soft form factors. The point is that the ratios
 $\zeta_\parallel^{K^*}/\zeta_\parallel^{\rho}$ and
$ \zeta_\perp^{K^*}/\zeta_\perp^{\rho} $ are more reliably calculable than
the form factors themselves.
\subsection{Input parameters}
To get the differential distributions numerically, some input parameters have to be specified.
For the calculation of the Wilson coefficients, the relevant parameters are chosen as \cite{PDG04}
\begin{equation}
M_W=80.425~\mbox{GeV}~,\hspace*{0.5cm} \sin^2 \theta_W=0.2312~, \hspace*{0.5cm}
\Lambda^{(5)}_{\overline{MS}}=217^{+25}_{-23}~\mbox{MeV}~,
\end{equation}
and $m_t^{pole}=(172.7 \pm 2.9)~\mbox{GeV}$, updated recently by the Tevatron
electroweak group \cite{CDFD0}.
Numerical values of the Wilson coefficients, evaluated at scale $\mu=m_b=4.8~\mbox{GeV}$,
 with the
three-loop running of $\alpha_s$ and the input parameters fixed at their
central values given above are shown in Table I.
Note that the NNLL formula for $C_9$ can be found, for example, in the
appendix of \cite{Beneke01}, while the relevant elements of three-loop
anomalous dimension matrix have been calculated recently in
\cite{Gambino03,Gorbahn05}.
\begin{table}[htb]
\begin{center}
\caption{The leading-logarithmic (LL) and next-to-leading-logarithmic (NLL)
Wilson coefficients evaluated at the scale $m_b=4.8~\mbox{GeV}$. For $C_{9,10}$,
they are also given in the NNLL order.}
\begin{tabular}{|c|c|c|c|c|c|c|}\hline
 ~ & LL & NLL & ~ & LL & NLL & NNLL\\ \hline
 $~~~\bar{C}_1$~~~ & ~~-0.2501~~ & ~~-0.1459~~ &
 ~~~$\bar{C}_6$~~~ & ~~-0.0316~~ & ~~-0.0388~~ & ~ \\ \hline
$\bar{C}_2$ & 1.1082 & 1.0561 & $C_7^{eff}$ & -0.3145 & -0.3054 & ~ \\ \hline
$\bar{C}_3$ & 0.0112 & 0.0116 & $C_8^{eff}$ & -0.1491 & -0.1678 & ~ \\ \hline
$\bar{C}_4$ &-0.0257 &-0.0337 & $   C_9   $ &  1.9919 &  4.1777 & ~~4.2120~~ \\ \hline
$\bar{C}_5$ & 0.0075 & 0.0097 & $   C_{10}$ &    0    & -4.5415 & -4.1958 \\ \hline
\end{tabular}
\end{center}
\end{table}

The CKM factor $\vert V_{ts} V_{tb}^* \vert \simeq (1-\lambda^2/2)
\vert V_{cb} \vert$ is estimated to be $0.0403 \pm 0.0020$ by taking
$\vert V_{cb} \vert = 0.0413 \pm 0.0021$ \cite{HFAG} and $\lambda=0.2226$.
For the B meson lifetimes, we use $\tau_{B^+}=1.643 ~\mbox{ps}$ and $
\tau_{B^0}=1.528~\mbox{ps}$ \cite{HFAG}. The pole mass $m_b$ is chosen to be
$4.8~\mbox{GeV}$. The ratio of the charm quark
mass over the b-quark mass is taken to be $m_c/m_b=0.29 \pm 0.02$. For the
matching scale from SCET$_I$ to SCET$_{II}$, we use $\mu_l=\sqrt{m_b \Lambda_h}
\simeq 1.5~\mbox{GeV}$.

The hadronic parameters for the decay $B \to K^* \ell^+\ell^-$ include decay
 constants,
light-cone distribution amplitudes (LCDAs) and the soft form
factors. The B meson decay constant can be estimated by QCD sum rules or lattice
calculations, here we take $f_B=(200 \pm 30)~\mbox{MeV}$. For the $K^\ast$ meson,
experimental measurements give \cite{PDG04} $f^\parallel_{K^\ast}=(217 \pm 5)
~\mbox{MeV}$ while the most recent light-cone sum rules (LCSRs)
estimate \cite{Ball0510} is
$f^\perp_{K^\ast}(1~\mbox{GeV})=(185 \pm 10)~\mbox{MeV}$. Note that
$f^\perp_{K^\ast}$ obeys the scale evolution equation $f^\perp_{K^\ast}(\mu)=
f^\perp_{K^\ast}(\mu_0)(\alpha_s(\mu)/\alpha_s(\mu_0))^{4/23}$.

The B meson LCDAs
enter into the decay amplitudes only in terms of the integrated quantities
$\lambda_{B,+}^{-1} $ and $ \lambda_{B,-}^{-1}(q^2) $ defined as
by the following integrals
\begin{equation}
\lambda_{B,+}^{-1} \equiv \int_0^\infty \! \frac{d\omega}{\omega}
\phi^B_+(\omega)~, \hspace*{1cm}
\lambda_{B,-}^{-1}(q^2)\equiv \int_0^\infty \! d\omega \frac{\phi^B_-(\omega)}
{\omega-q^2/m_b-i\epsilon}~.
\end{equation}
Therefore, it is not necessary to know the details about the shape of
$\phi^B_+(\omega)$. The most recent estimate gives \cite{Neubert05B}
$\lambda_{B,+}^{-1}=(1.86 \pm 0.34)~\mbox{GeV}^{-1}$ at the scale
$\mu=1.5 ~\mbox{GeV}$. However, $\lambda_{B,-}^{-1}(q^2)$ does require the
knowledge of $\phi^B_-(\omega)$, about which we know very little. Fortunately,
$\lambda_{B,-}^{-1}(q^2)$ only appears in the annihilation term which plays
numerically a minor role in the $B \to K^\ast \ell^+ \ell^-$ decay. To be definite, we
adopt a simple model function \cite{Neubert97}
$\phi^B_-(\omega)=\omega_0^{-1}e^{-\omega/\omega_0}$ with
$\omega_0^{-1} \simeq 3~\mbox{GeV}^{-1}$.

The $K^\ast$ meson LCDAs may be
expanded in terms of Gegenbauer polynomials:
\begin{equation}
\phi_{K^\ast}^{\perp,\parallel}(u,\mu)=6u(1-u)\left [ 1 + \sum_{n=1}^{\infty}
a_n^{\perp,\parallel} (\mu) C_n^{3/2}(2u-1) \right ]~.
\end{equation}
However, the coefficients $a_n$ are largely unknown. Following \cite{Ball05},
we shall ignore the terms $a_n^{\perp,\parallel}~(n>2)$.
For $a_{1,2}$, we omit their scale dependence and estimate in a conservative
manner:
$a_1^{\perp,\parallel}=0.1 \pm 0.1$, $a_2^{\perp,\parallel}=0.1 \pm 0.1$.
We note that recently the first Gegenbauer moment of the
$K^\ast$ meson has been revisited in LCSRs \cite{Ball0510} which
gives smaller uncertainties.

There are only two independent $B \to K^\ast$ form factors in SCET, namely
$\zeta_\perp(q^2)$ and $\zeta_\parallel(q^2)$. They are related to the full QCD
form factors as discussed in \cite{Neubert04}. The current knowledge of these
form factors is fragmentary. For instance, $\zeta_\perp$
may be extracted from $V^{B\to K^\ast}$ \cite{Neubert05}:
\begin{equation}
\zeta_\perp(q^2)=\frac{\zeta_\perp(0)}{r_1^V+r_2^V}
\left ( \frac{r_1^V}{1-q^2/m_V^2} + \frac{r_2^V}{1-q^2/m_{Vfit}^2} \right )~,
\end{equation}
with $r_1^V=0.923$, $r_2^V=-0.511$,
$m_V=5.32~\mbox{GeV}$ and $m_{Vfit}^2=49.40~\mbox{GeV}^2$.
Note that the $q^2$-dependence above is the same as
that of $V^{B\to K^\ast}(q^2)$, calculated
in \cite{Ball05} using LCSRs. However, analyses of the radiative B decays $B \to K^\ast \gamma$
\cite{Ali01,Beneke01,Buchalla01,Neubert05}, $B \to \rho \gamma$
\cite{Ali01,Buchalla01} and the semi-leptonic B decay $B \to \rho \ell \nu$
\cite{BaBarrho} imply that the LCSRs overestimate the
$B \to V$ form factors significantly. We use the radiative $B \to K^\ast
\gamma$ decay,
 which has been measured quite precisely
\cite{HFAG}: ${\cal B}(B^0 \to K^{\ast0} \gamma)=(4.01 \pm 0.20) \times
10^{-5}$, to normalize the
soft form factor at $q^2=0$.
In SCET, it is straightforward to get the decay amplitude of $B \to K^\ast \gamma$ from the $B \to
K^\ast_\perp \ell^+ \ell^-$ decay, by taking the limit $q^2 \to 0$. Then, using the input parameters
from Table II, we obtain $\zeta_\perp(0)=0.32 \pm 0.02$. Here the error is mainly from the CKM factor
$V_{ts}V^*_{tb}$ and the experimental uncertainty of the branching ratio
 ${\cal B}(B^0 \to K^{\ast0} \gamma)$.
This estimate is consistent with the result of Ref. \cite{Neubert05}, but
significantly smaller
than the number $0.40 \pm 0.04$ we get from LCSRs.
In our numerical analysis, we will choose the value $\zeta_\perp(0)=0.32 \pm 0.02$
determined from the radiative B decays, but assume that the $q^2$-dependence of
 $\zeta_\perp(q^2)$ can be reliably obtained from the LCSRs.

 For the longitudinal soft form factor $\zeta_\parallel$,
unfortunately there is no quantitative determination from the existing experiments,
though this may change in the future with good quality data available on the
decay $B \to \rho \ell \nu_\ell$. Using helicity analysis, one can extract
$\zeta^{\rho}_\parallel(q^2)$; combined with estimates of the
SU(3)-breaking one may determine $\zeta^{K^*}_\parallel(q^2)$. Not having this
experimental information at hand, one may
extract $\zeta_\parallel(q^2)$ from the full QCD form factor
$A_0^{B\to K^\ast}(q^2)$:
\begin{equation}
\begin{aligned}
A_0^{B\to K^\ast}(q^2)&=
\left [ 1-\frac{\alpha_s(m_b) C_F}{4\pi} \left ( 2 \ln^2 [1-s]
-\frac{2}{s}\ln [1-s] + 2~Li_2[s] +4 + \frac{\pi^2}{12} \right ) \right ]
\zeta_\parallel (q^2)\\
& \hspace*{-1.1cm}
- \frac{1}{4(1-s)}f_B\phi_+^B \otimes f_{K^\ast}^\parallel
\phi_{K^\ast}^\parallel \otimes {\cal J}_\parallel \otimes
\left ( \frac{2E}{\mu_h} \right )^{a(\mu_h,\mu_l)}
e^{S(\mu_h,\mu_l)} \int_0^1 \! dy~U_\parallel (v,y,\mu_h,\mu_l)~\end{aligned}
\end{equation}
with $s=q^2/m_B^2$. LCSRs estimate \cite{Ball05}
$A_0^{B\to K^\ast}(0)=0.374 \pm 0.043$ with the $q^2$-dependence
\begin{equation}
A_0^{B\to K^\ast}(q^2)=\frac{1.364}{1-q^2/m_B^2}-\frac{0.990}{1-q^2/36.78 \mbox{GeV}^2}~.
\end{equation}
From which we get $\zeta_\parallel(0)=0.40 \pm 0.05$, using the input
parameters discussed above and/or listed in Table II. Its $q^2$-dependence is
drawn in Fig. 4.
\begin{table}[tb]
\begin{center}
\caption{Numerical values of the input parameters and their uncertainties used
in the phenomenological study.}
\begin{tabular}{|ll|ll|}\hline
$M_W$ & $80.425~\mbox{GeV}$ & $\sin^2 \theta_W$ & $0.2312$ \\ \hline
$m_t^{pole}$& $(172.7 \pm 2.9)~\mbox{GeV} $ & $\Lambda^{(5)}_{\overline{MS}}$
& $(217^{+25}_{-23})~\mbox{MeV}$ \\ \hline
$\vert V_{ts} V^*_{tb} \vert$ & $(40.3 \pm 2.0) \times 10^{-3}$ &
$\alpha_{em}(m_b)$ & $1/133$ \\ \hline
$m_B$ & $5.279~\mbox{GeV}$ & $m_b^{pole}$ & $4.8~\mbox{GeV}$
\\ \hline
$\tau_{B^+}$  & $1.643~$ps &
$\tau_{B^0}$ \hspace*{2.2cm} &  $1.528~$ps  \\ \hline
$m_c/m_b$ & $0.29 \pm 0.02$ & $\mu_l$ & $1.5~\mbox{GeV}$ \\ \hline
$\lambda^{-1}_{B,+}(1.5~\mbox{GeV})$\hspace*{0.2cm} &
$(1.86 \pm  0.34)~\mbox{GeV}^{-1}$ & $f_B$ & $(200 \pm 30)~\mbox{MeV}$ \\ \hline
$\zeta_\perp(0)$ & $0.32 \pm 0.02$ & $\zeta_\parallel(0)$ & $0.40 \pm 0.05$
\\ \hline
$f_{K^\ast}^\perp (1~\mbox{GeV})$ & $(185 \pm 10)~\mbox{MeV}$ &
$f_{K^\ast}^\parallel$ & $(217 \pm 5)~\mbox{MeV}$ \\ \hline
$a_1^{\perp,\parallel}$ & $0.1 \pm 0.1$ &
$a_2^{\perp,\parallel}$ & $0.1 \pm 0.1$ \\ \hline
\end{tabular}
\end{center}
\end{table}

Alternatively, $\zeta_\parallel(q^2)$ may also be determined from the
following relation,
\begin{equation}
\begin{aligned}
\frac{E m_B (V-A_2)^{B \to K^*}(q^2)}{m_{K^\ast}(m_B+m_{K^\ast})} &=
\left [ 1-\frac{\alpha_s(m_b) C_F}{4\pi} \left ( 2 \ln^2 [1-s]
-2\ln [1-s] + 2~Li_2[s] +6 + \frac{\pi^2}{12} \right ) \right ]
\zeta_\parallel (q^2)\\
& \hspace*{-2.1cm}
- \frac{1-2s}{4(1-s)}f_B\phi_+^B \otimes f_{K^\ast}^\parallel
\phi_{K^\ast}^\parallel \otimes {\cal J}_\parallel \otimes
\left ( \frac{2E}{\mu_h} \right )^{a(\mu_h,\mu_l)}
e^{S(\mu_h,\mu_l)} \int_0^1 \! dy~U_\parallel (v,y,\mu_h,\mu_l)~.
\end{aligned}
\end{equation}
With the input $V^{B\to K^\ast}(0)-A_2^{B\to K^\ast}(0)=0.152 \pm 0.057$ from
LCSRs, we obtain $\zeta_\parallel(0)=0.42 \pm 0.16$, which
agrees with the range extracted from $A_0^{B\to K^\ast}$.
We will use $\zeta_\parallel(0)=0.40 \pm 0.05$, obtained from its
relation to the full form factor
$A_0^{B\to K^\ast}$ and the LCSR, as discussed above. Fig. 4 shows the $q^2$-dependence of both soft form
factors $\zeta_{\perp,\parallel}(q^2)$. However, since the analysis of the semileptonic decay
$B \to \rho \ell \nu$ \cite{BaBarrho} suggests that both the transverse and longitudinal form
factors might be overestimated by LCSRs, we will also consider, as an
illustration of the non-perturbative uncertainties, the value
$\zeta_\parallel(0)=\zeta_\perp(0)=0.32$
 with all the other parameters taken at their central values.

\begin{figure}[tb]
\begin{center}
\unitlength 1mm
\begin{picture}(80,50)
\put(0,0){\includegraphics[width=0.5\textwidth]{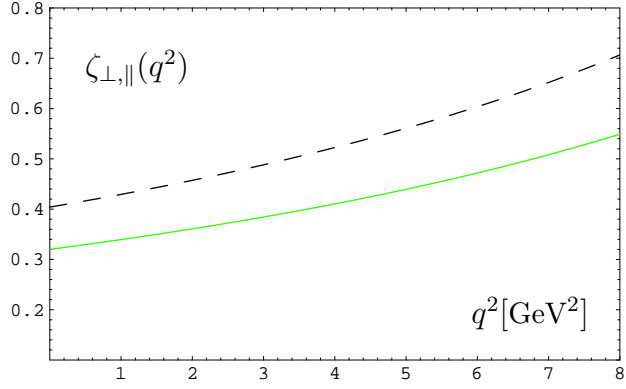}}
\end{picture}
\vspace*{-0.2cm} \caption{The $q^2$-dependence of the soft form
factors $\zeta_{\perp,\parallel}(q^2)$. The solid curve represents
$\zeta_\perp(q^2)$, while the dashed curve represents $\zeta_\parallel(q^2)$. We
have rescaled the transverse form factor at $q^2=0$, to be consistent with the experimental
measurements of the $B \to K^* \gamma$ decay rate.}
\end{center}
\end{figure}

\subsection{Numerical solution of the SCET$_I$ evolution functions}
As we discussed in Sect. II.D, the B-type matching coefficients $C_i^B$ should
be run from the scale $\mu_h=4.8~\mbox{GeV}$ down to $\mu_l=1.5~\mbox{GeV}$,
with the evolution kernel $\widetilde{U}_\Gamma (E,u,\mu_h,\mu)$ obeying the
integro-differential equation (\ref{kernel}). To solve this equation numerically,
it is more convenient to define the following evolution functions,
\begin{equation}
\begin{aligned}
\widetilde{U}_\Gamma^{(a)}(E,u,\mu_h,\mu)&=\int_0^1 \! dv
U_\Gamma(u,v,\mu_h,\mu)~, \\
\widetilde{U}_\Gamma^{(b)}(E,u,\mu_h,\mu)&=\frac{u+\hat{s}-u\hat{s}}{1-u}
\int_0^1 \! dv U_\Gamma(u,v,\mu_h,\mu) \frac{1-v}{v+\hat{s}-v\hat{s}}~, \\
\widetilde{U}_\Gamma^{(c)}(E,u,\mu_h,\mu)&=\int_0^1 \! dv
U_\Gamma(u,v,\mu_h,\mu) \frac{F_{16}^\Gamma(v,\hat{s},m_c^2/m_b^2)}
{F_{16}^\Gamma(u,\hat{s},m_c^2/m_b^2)}~,
\end{aligned}
\end{equation}
where $\Gamma=\perp,\parallel$ and the functions
$F_{16}^{\perp,\parallel}(u,\hat{s},m_c^2/m_b^2)$ are defined in
Eqs.~(\ref{F16perp}) and (\ref{F16para}). Note that at
the quark level, the $K^\ast$ meson
energy is related to $\hat{s}$ by $E=m_b(1-\hat{s})/2$ in the rest frame
of the $b$-quark.
With such definitions, the above evolution functions are normalized to one at
the scale $\mu_h$:
$\widetilde{U}_\Gamma^{(a,b,c)}(E,u,\mu_h,\mu_h)=1$,
and the QCD parameter $\Lambda^{(5)}_{\overline{MS}}$ would be the only input
for their numerical evaluations.
The matching coefficients $C_j^B$ at scale $\mu_l$ can then be written as
\begin{equation}
\Delta_i C_j^B(E,u,\mu_l)=\left (\frac{2E}{\mu_h} \right )^{a(\mu_h,\mu_l)}
e^{S(\mu_h,\mu_l)}
\widetilde{U}_\Gamma^{(a,b,c)}(E,u,\mu_h,\mu_l) \Delta_i C_j^B(E,u,\mu_h)~,
\end{equation}
where we should use the superscript $(a)$ for $\Delta_{7,9,10}C_j^B$, the
superscript $(b)$ for $\Delta_8 C_j^B$ and the superscript $(c)$ for
$\Delta_{16} C_j^B$. For the subscript $\Gamma$, one should use $\Gamma=\perp$
for $j=1,3$ and $\Gamma=\parallel$ for $j=2,4$, which is the same as the
convention of Eq.~(\ref{evolution}). Note that for the evolution of
$\Delta_{16} C_j^B$, we have taken into account the fact that
the term $F^\Gamma_{16}(u,\hat{s},m_c^2/m_b^2)$ is dominant due to the
large Wilson coefficient $\bar{C}_2$.

It is then straightforward to get the following evolution equations
\begin{equation}
\begin{aligned}
\frac{d\widetilde{U}^{(a)}_\Gamma (E,u,\mu_h,\mu)}{d\ln\mu}&=\int_0^1 \!
dy~ y V_\Gamma (y,u)\widetilde{U}^{(a)}_\Gamma (E,y,\mu_h,\mu) +
\omega(u)\widetilde{U}^{(a)}_\Gamma (E,u,\mu_h,\mu)~, \\
\frac{d\widetilde{U}^{(b)}_\Gamma (E,u,\mu_h,\mu)}{d\ln\mu}&=\int_0^1 \!
dy~ y V_\Gamma (y,u)\frac{(1-y)(u+(1-u)\hat{s})}{(1-u)(y+(1-y)\hat{s})}
\widetilde{U}^{(b)}_\Gamma (E,y,\mu_h,\mu) + \\
& \hspace*{1cm} \omega(u)\widetilde{U}^{(b)}_\Gamma (E,u,\mu_h,\mu)~, \\
\frac{d\widetilde{U}^{(c)}_\Gamma (E,u,\mu_h,\mu)}{d\ln\mu}&=\int_0^1 \!
dy~ y V_\Gamma (y,u)\frac{F_{16}^\Gamma(y,\hat{s},m_c^2/m_b^2)}
{F_{16}^\Gamma(u,\hat{s},m_c^2/m_b^2)}\widetilde{U}^{(c)}_\Gamma (E,y,\mu_h,\mu) + \\
& \hspace*{1cm}\omega(u)\widetilde{U}^{(c)}_\Gamma (E,u,\mu_h,\mu)~.
\end{aligned}
\end{equation}
\begin{figure}[tb]
\begin{center}
\unitlength 1mm
\begin{picture}(160,120)
\put(0,0){\includegraphics[width=0.9\textwidth]{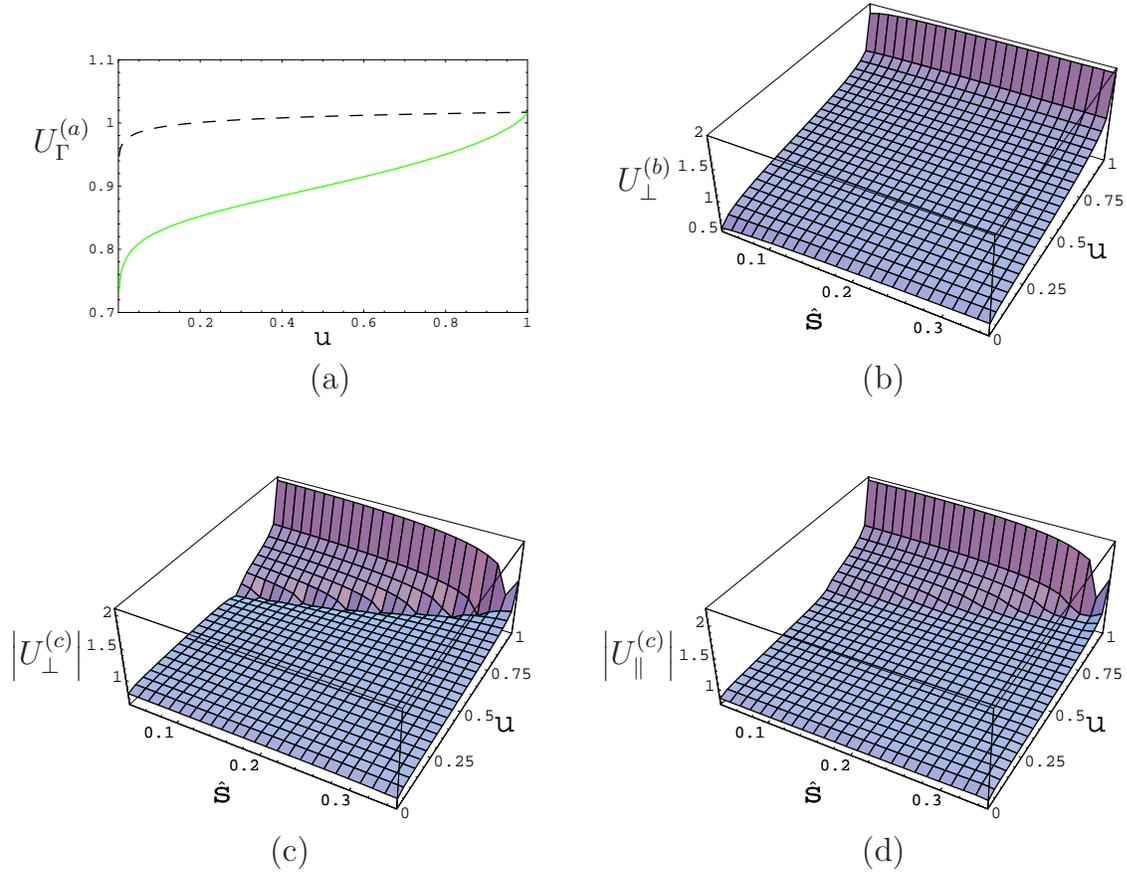}}
\end{picture}
\vspace*{-0.2cm} \caption{Numerical values of the functions
$\widetilde{U}^{(a,b,c)}_\Gamma (E,u,\mu_h,\mu_l)$, evolved from
$\mu_h=4.8~\mbox{GeV}$ down to $\mu_l=1.5~\mbox{GeV}$, the relevant
parameters are taken at their central values. For the upper-left plot,
the solid line denotes $\widetilde{U}_\perp^{(a)}$ while the dashed line denotes
$\widetilde{U}_\parallel^{(a)}$. For the lower plots, since
$\widetilde{U}_\Gamma^{(c)}$ are complex functions, we only show
their absolute values. }
\end{center}
\end{figure}
To get the numerical solutions of the above integro-differential equations, we
will perform the scale evolution in one hundred discrete steps. While from the
scale $\mu_n$ to $\mu_{n+1}$, the convolution integral is evaluated for three
hundred different values and discrete $\hat{s}$ values of
$\delta \hat{s}=0.01$ in the interval $\hat{s}\in[0.04,0.35]$.
The function $\widetilde{U}_\Gamma (E,u,\mu_h,\mu_{n+1})$ is obtained from
a fit to these values. Taking $\Lambda_{\overline{MS}}^{(5)}=217~\mbox{MeV}$, the
numerical results of these evolution functions are shown in Fig. 5. Note that
the function $\widetilde{U}_\Gamma^{(a)}(E,u,\mu_h,\mu)$ actually does not
depend on the energy $E$, as shown in Fig. (5a). In fact, it is just
the same function as $U_\Gamma(u,\mu_h,\mu)$ defined in Eq.~(5.23) by
Neubert {\it et al.} \cite{Neubert04}. The function
$\widetilde{U}_\parallel^{(b)}$ is not shown in Fig. 5, since it does not
enter into the decay amplitude at the one-loop level, due to $\Delta_8 C_2^B=0$.
While for the complex functions $\widetilde{U}_\Gamma^{(c)}$, only the absolute
values of the functions are plotted.

\subsection{The dilepton invariant mass spectrum and the forward-backward asymmetry}
Experimentally, the dilepton invariant mass spectrum and the
forward-backward (FB) asymmetry are the observables of principal
interest. Their theoretical expressions in SCET
can be easily derived from Eq.~(\ref{decayamplitude}):

\begin{equation}
\begin{aligned}
\frac{d Br}{d q^2} &= \tau_B \frac{G_F^2 \vert
V_{ts}^\ast V_{tb} \vert^2}{128\pi^3} \left
(\frac{\alpha_{em}}{4\pi} \right )^2 m_B^3 \vert \lambda_{K^\ast}
\vert (1-\frac{q^2}{m_B^2})^2 \times \\
& \left \{ \frac{16}{3}\zeta_\perp^2 \frac{q^2}{m_B^2} (\vert
{\cal C}_{9}^\perp \vert^2 + ({\cal C}_{10}^\perp)^2 )
+\frac{4}{3}\zeta_\parallel^2(\vert {\cal C}_{9}^\parallel \vert^2 + ({\cal
C}_{10}^\parallel)^2 ) \right \}~,
\end{aligned}
\end{equation}
\begin{equation}\label{DAFB}
\begin{aligned}
\frac{d A_{FB}}{d q^2}&=\frac{1}{d \Gamma / d q^2} \left (
\int_0^1 \! d \cos \theta \frac{d^2 \Gamma}{d q^2 d\cos \theta} -
\int_{-1}^0 \! d \cos \theta \frac{d^2 \Gamma}{d q^2 d\cos \theta} \right ) \\
&=\frac{-6 (q^2/m_B^2) \zeta_\perp^2 Re({\cal C}_{9}^\perp)
{\cal C}_{10}^\perp }{4(q^2/m_B^2) \zeta_\perp^2 (\vert
{\cal C}_{9}^\perp \vert^2 + ({\cal C}_{10}^\perp)^2 )
+\zeta_\parallel^2(\vert {\cal C}_{9}^\parallel \vert^2 + ({\cal
C}_{10}^\parallel)^2 ) }~.
\end{aligned}
\end{equation}
With the input parameters listed in Table II, the decay spectrum and the FB
asymmetry are shown in Fig. 6 and Fig. 7, respectively.
In our calculation we have dropped the small isospin-breaking effects,
which come from the annihilation diagrams, and take the spectator quark as the
down quark in Eqs. (\ref{anni1}, \ref{anni2}). To estimate the
residual scale dependence, we vary the QCD matching scale $\mu_h$ by a
factor $\sqrt{2}$ around the default value $\mu_h=m_b$. Note that the soft
form factors $\zeta_{\perp,\parallel}(q^2)$ defined in SCET are actually scale
dependent, which effect has been taken into account in our error analysis.
\begin{figure}[tb]
\begin{center}
\unitlength 1mm
\begin{picture}(160,53)
\put(0,0){\includegraphics[width=0.9\textwidth]{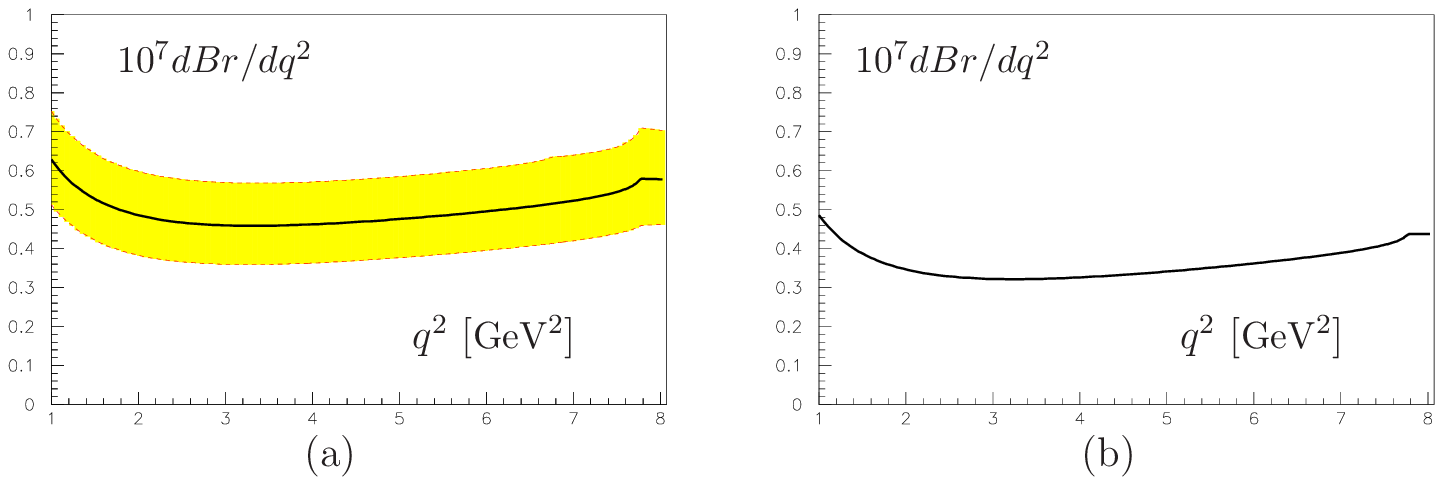}}
\end{picture}
\vspace*{-0.2cm} \caption{The differential branching ratio
$d {\cal B}(B^0 \to K^{\ast 0} \ell^+ \ell^-)/d q^2$ in the range
$1~{\rm GeV}^2 \leq q^2 \leq 8~~{\rm GeV}^2$. In the left plot,
the solid line denotes the theoretical prediction
with the input parameters taken at their central values, while the gray
area between two dashed lines reflects the uncertainties from input parameters
and scale dependence. In the right plot, the soft form factors are normalized as
$\zeta_\parallel(0)=\zeta_\perp(0)=0.32$, while all the other parameters
 are chosen at their central values.}
\end{center}
\end{figure}

\begin{figure}[tb]
\begin{center}
\unitlength 1mm
\begin{picture}(80,50)
\put(0,0){\includegraphics[width=0.5\textwidth]{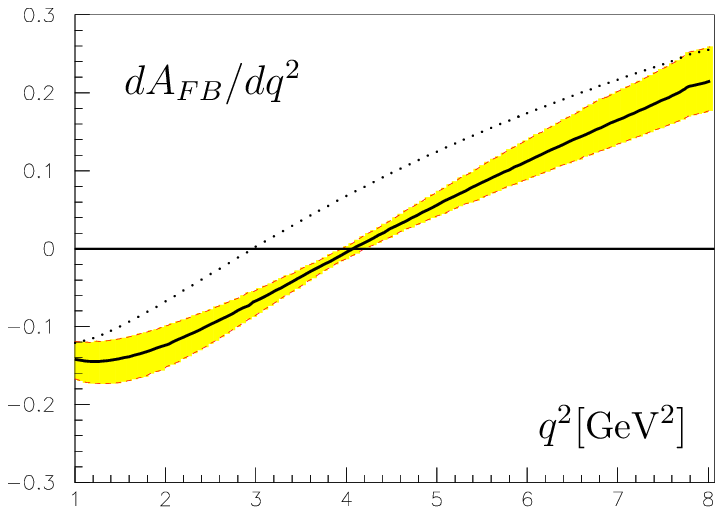}}
\end{picture}
\vspace*{-0.2cm} \caption{The differential spectrum of the forward-backward
 asymmetry $d A_{FB}(B \to K^\ast \ell^+ \ell^-)/d q^2$ in the range
$1~{\rm GeV}^2 \leq q^2 \leq 8~~{\rm GeV}^2$. Here the solid
line denotes the theoretical prediction
with the input parameters taken at their central values, while the gray
band between two dashed lines reflects the uncertainties from input parameters
and scale dependence. The dotted line represents the LO predictions, obtained by
dropping the $O(\alpha_s)$ corrections.}
\end{center}
\end{figure}

Restricting to the integrated branching ratio of $B \to K^\ast \ell^+ \ell^-$
in the range $1~\mbox{GeV}^2 \leq q^2 \leq 7~\mbox{GeV}^2$, where the SCET
method should work, we obtain
\begin{equation}
\int \limits_{1\mbox{\scriptsize ~GeV}^2}^{7\mbox{\scriptsize ~GeV}^2} d q^2
\frac{d Br(B^+ \to K^{\ast +} \ell^+ \ell^-)}{dq^2}=(2.92^{+0.57}_{-0.50}
\vert_{\zeta_\parallel}~^{+0.30}_{-0.28} \vert_{\mbox{\scriptsize CKM}}
~^{+0.18}_{-0.20})\times 10^{-7}~.
\end{equation}
Here we have isolated the uncertainties from the soft form factor
$\zeta_\parallel$ and the CKM factor $\vert V_{ts}^* V_{tb} \vert$.
The last error reflects the uncertainty
due to the variation of the other input parameters and the residual scale
dependence. If the smaller value for the longitudinal form factor $\zeta_\parallel(0)=
0.32$ is used, as shown in Fig. (6b), the central value of the branching ratio
is reduced  to $2.11 \times 10^{-7}$.
For $B^0$ decay, the branching ratio is about $7\%$ lower due to the
lifetime difference:
\begin{equation}
\int \limits_{1\mbox{\scriptsize ~GeV}^2}^{7\mbox{\scriptsize ~GeV}^2} d q^2
\frac{d Br(B^0 \to K^{\ast 0} \ell^+ \ell^-)}{dq^2}=(2.72^{+0.53}_{-0.47}
\vert_{\zeta_\parallel}~^{+0.28}_{-0.26} \vert_{\mbox{\scriptsize CKM}}
~^{+0.17}_{-0.19})\times 10^{-7}~.
\end{equation}

To compare with the current experimental observations, it was proposed in
Ref.~\cite{Beneke01} to consider the integrated branching ratio over the range
$4~\mbox{GeV}^2 \le q^2 \le 6~\mbox{GeV}^2$, for which we get
$(0.92 ^{+0.21}_{-0.19}) \times 10^{-7}$. This is smaller than the number
$(1.2 \pm 0.4) \times 10^{-7}$ obtained in Ref. \cite{Beneke01}, which is mainly
due to the fact that the most recent LCSRs estimation \cite{Ball05} prefers
the form factor $A_0^{B \to K^\ast}$ to be smaller. Experimentally one of the
Belle observations \cite{Belle} of our interest is
\begin{equation}
\int \limits_{4\mbox{\scriptsize ~GeV}^2}^{8\mbox{\scriptsize ~GeV}^2} d q^2
\frac{d Br(B \to K^\ast \ell^+ \ell^-)}{dq^2}=(4.8^{+1.4}_{-1.2}
\vert_{\mbox{\scriptsize stat.}}\pm 0.3 \vert_{\mbox{\scriptsize syst.}}
\pm 0.3 \vert_{\mbox{\scriptsize model}})\times 10^{-7}~,
\end{equation}
for which we predict $(1.94^{+0.44}_{-0.40}) \times 10^{-7}$. This is smaller
than the published Belle data by a factor of about 2.5. But at this
stage, it is still too early to conclude that one should change some theoretical
input significantly to be consistent with the experimental data. For
instance, the BaBar collaboration measures the total branching ratio of
$B \to K^\ast \ell^+ \ell^-$ to be \cite{BaBar}
$(7.8^{+1.9}_{-1.7}\pm 1.2) \times 10^{-7}$,
which is about twice smaller than the Belle observation \cite{Belle}
$(16.5^{+2.3}_{-2.2}\pm 0.9 \pm 0.4) \times 10^{-7}$. This implies that, if
finally the total branching ratio of $B \to K^\ast \ell^+ \ell^-$ is found to
be closer to the BaBar result, the partially integrated branching ratio over the
range $4~\mbox{GeV}^2 \le q^2 \le 8~\mbox{GeV}^2$ could be lowered to around
$2.3 \times 10^{-7}$, which is consistent with our estimate
$(1.94^{+0.44}_{-0.40}) \times 10^{-7}$ within the stated errors. We look
forward to experimental analyses from BaBar and Belle based on their
high statistic data.

One of the most interesting observables in the decay $B \to K^\ast \ell^+ \ell^-$ is
the location, $q_0^2$,  where the FB asymmetry vanishes. It was first noticed in
the context of form factor models in \cite{Burdman} and later demonstrated in
\cite{Ali00},
using the symmetries of the effective theory in the large-energy limit, that
the value of $q_0^2$ is
almost free of hadronic uncertainties at leading order. From Eq. (\ref{DAFB}), it is easy to see
that the location of the vanishing FB asymmetry is determined by
$Re({\cal C}_{9}^\perp)=0$.
 At the leading order, this leads to the equation
$C_9+C_7^{eff}+Re(Y(q_0^2))=0$.
Including the order $\alpha_s$ corrections, our analysis estimates the
zero-point of the FB asymmetry to be
\begin{equation}
q^2_0=(4.07^{+0.16}_{-0.13})~ \mbox{GeV}^2~,
\end{equation}
of which the scale-related uncertainty is
 $\Delta(q_0^2)_{\rm scale}=^{+0.08}_{-0.05}$
 GeV$^2$ for the range
 $m_b/2 \leq \mu_h \leq 2 m_b$ together with the jet function scale
$\mu_l=\sqrt{\mu_h \times 0.5~\mbox{GeV}}$, as used in the
paper by  Beneke et al.~\cite{Beneke01}. Since no reliable estimates of the
power corrections in $1/m_b$ are available, we should compare our results
with the one given in Eq.~(74) of~\cite{Beneke01}, also obtained in the
absence of $1/m_b$ corrections:
$ q^2_0=(4.39^{+0.38}_{-0.35})~ \mbox{GeV}^2$. Of this the largest single
uncertainty (about $\pm 0.25~ \mbox{GeV}^2 $) is attributed to the scale
dependence. While our central value for $q_0^2$ is similar to theirs, with the
differences reflecting the different input values, the scale
dependence in our analysis is significantly smaller than that of
\cite{Beneke01}. This improved theoretical precision on $q_0^2$ requires a
detailed discussion to which we now concentrate in the rest of this section.

As already stated in the introduction, the expressions  for the differential distributions in the decay
$B \to K^* \ell^+\ell^-$ derived here and in \cite{Beneke01} are
similar except for the definitions of the soft form factors
and the additional step of the SCET logarithmic resummation incorporated in our
paper. This resummation has also been derived in the existing
literature~\cite{Neubert05,Neubert04,Beneke:2005gs}. However, its effect
on the scale-dependence of $q_0^2$ has not been studied
in sufficient detail. With the SCET form factors $\zeta_\perp(q^2,\mu)$
and  $\zeta_\parallel(q^2,\mu)$ defined in Eq.~(33) here, which are
scale-dependent quantities, and neglecting the resummation effects
consistently in both the decays
 $B \to K^* \ell^+ \ell^-$ and $B \to K^* \gamma$, the scale uncertainty 
is increased, with $q_0^2=4.12^{+0.17}_{-0.07}$ GeV$^2$.
We draw two inferences from this numerical study: (i)
Incorporating the SCET
logarithmic resummation helps in the reduction of scale dependence
in $q_0^2$,
(ii) $\Delta (q_0^2)_{\rm scale}=~^{+0.17}_{-0.07}$, obtained by dropping the
resummation effects is still significantly smaller (by a factor 2) compared to
the corresponding uncertainty
$\Delta(q_0^2)_{\rm scale}=~\pm 0.25$ GeV$^2$ calculated in
 Ref.~\cite{Beneke01}.
This deifference, as argued below, is to be traced back to 
the different definitions of the soft form factors used by us
for the SCET currents and the corresponding quantities employed by
Beneke et al.~\cite{Beneke01} in the QCD factorization approach.
The results in Ref.~\cite{Beneke01} are, however, formally equivalent
to the so-called "physical form factor'' (PFF)
scheme  in SCET, as discussed subsequently by Beneke and
Yang~\cite{Beneke:2005gs}. Thus, the scale dependence of the 
distributions in $B \to K^* \ell^+ \ell^-$, in particular of $q_0^2$, is
related also to the definitions (or scheme dependence) of the form factors 
in effective theories. The PFF-scheme is one such choice, but this choice is
by no means unique. 

Concentrating on the transverse
form factor, relevant for $q_0^2$ of the FB asymmetry, in the PFF
scheme, the corresponding SCET$_I$
form factor $\zeta_\perp^P$ (where we have now added a suffix $P$ for this
scheme) is defined as
\begin{equation}
\zeta_\perp^P\equiv \frac{m_B}{m_B+m_{K^*}}V~,
\end{equation}
where $V$ is one of the physical form factors in the decay $B \to K^*
\ell^+\ell^-$ in full QCD. In contrast, in our paper, the soft SCET form
factors are defined in Eq.
(\ref{SCETff}). These two definitions can be related to each
other by $\zeta_\perp^P=\widetilde{C}_3 \zeta_\perp$, where the expression
for the perturbative QCD coefficient $\widetilde{C}_3$ is given below
 ($\widetilde{C}_3$ is called $C_V^{(A0)1}$ in \cite{Yang04}). Since the
decay amplitude should be independent on how one defines the soft form
factors, one must have
\begin{equation}
{\cal C}_9^{\perp P} \zeta_\perp^P \equiv {\cal C}_9^\perp \zeta_\perp
\Longrightarrow
{\cal C}_9^{\perp P}= {\cal C}_9^\perp / \widetilde{C}_3~,
\end{equation}
 Since
$\widetilde{C}_3=1+{\cal O}(\alpha_s)$, by expanding
${\cal C}_{9}^\perp/\widetilde{C}_3$ to order $\alpha_s$, one
obtains
\begin{eqnarray}
{\cal C}_9^{\perp P}&=&\frac{{\cal C}_9^\perp}{ 1-(1-\widetilde{C}_3)}
\nonumber \\
&\simeq&\frac{2\pi}{\alpha_{em}} \left ( C_1^A +\frac{\alpha_{em}}{2\pi}
(1 - \widetilde{C}_3)(\frac{2}{\hat{s}}C_7^{eff}+C_9^{eff}) +
\frac{m_B}{4}\frac{f_B \phi_+^B \otimes
f^\perp_{K^\ast}\phi^\perp_{K^\ast} \otimes {\cal J}_\perp \otimes
C_1^B}{\zeta^P_\perp} \right ) \nonumber \\
&=&C_9^{eff}+\frac{2}{\hat{s}}C_{7}^{eff}
\left ( 1+\frac{C_F\alpha_s}{4\pi}\left [
4 \ln\frac{m_b^2}{\mu^2}-4+\frac{1-\hat{s}}{\hat{s}}
\ln ( 1-\hat{s})  \right ] \right ) + ...~,
\end{eqnarray}
which agrees with the expression for ${\cal C}_9^{\perp P}$ in Eq.(40) of
\cite{Beneke01} (called $C_{9,\perp}(q^2)$ there). We recall that to determine $q_0^2$, we solve the equation
${\rm Re}~{\cal C}_{9}^\perp=0$, where now the quantity ${\cal C}_{9}^\perp$
is defined as follows
\begin{equation}
{\cal C}_{9}^\perp=\widetilde{C}_3(\mu)
C_9^{eff}+\frac{2}{\hat{s}} C_{7}^{eff}\frac{{\overline m}_b}{m_b}
\widetilde{C}_9(\mu)+...~,
\end{equation}
with the QCD coefficients \cite{Bauer2} ($\widetilde{C}_9$ is called
$C_T^{(A0)2}$ in \cite{Yang04})
\begin{eqnarray}
\widetilde{C}_3&=&1-\frac{\alpha_s C_F}{4\pi}\left [ 2\ln^2 \left (
\frac{\mu}{m_b} \right ) - (4\ln(1-\hat{s})-5)\ln \left (
\frac{\mu}{m_b} \right ) \right . \nonumber \\
&&\left . +2\ln^2(1-\hat{s})+2 \mbox{Li}_2(\hat{s})+\frac{\pi^2}{12}+
\left ( \frac{1}{\hat{s}}-3 \right ) \ln (1-\hat{s})+6 \right ]~,
\nonumber \\
\widetilde{C}_9&=&1-\frac{\alpha_s C_F}{4\pi}\left [ 2\ln^2 \left (
\frac{\mu}{m_b} \right ) - (4\ln(1-\hat{s})-7)\ln \left (
\frac{\mu}{m_b} \right ) \right . \nonumber \\
&&\left . +2\ln^2(1-\hat{s})-2\ln(1-\hat{s})+2
\mbox{Li}_2(\hat{s})+\frac{\pi^2}{12}+6 \right ]~.
\end{eqnarray}
The ellipses above denote the terms which are the same for ${\cal C}_9^{\perp P}$
and ${\cal C}_9^{\perp}$.
The functions multiplying the effective Wilson coefficients
 $C_9^{eff}$ and $C_{7}^{eff}$
appearing in ${\cal C}_9^{\perp P}$ and ${\cal C}_{9}^\perp$ in Eqs.~(64)
and (65), respectively, lead to
different scale-dependence for $q_0^2$.

 Our result for $q_0^2$ using the SCET
form factors has been given above in Eq.~(61) with the scale-dependent
 uncertainty
$\Delta(q_0^2)_{\rm scale}=^{+0.08}_{-0.05}$ GeV$^2$ . Note that we have 
considered in a correlated way the scale-dependence of $\zeta_\perp(\mu,q^2)$ in 
our analysis. To illustrate this, we use the experimental data on
the branching ratio of  $B \to K^* \gamma$ and the central values of the
other input parameters given in Table II, which yields the following
scale-dependence of the relevant form factor: $\zeta_\perp(0,\mu=2m_b)=0.34$ and
$\zeta_\perp(0,\mu=m_b/2)=0.30$. In solving the equation
 $Re[{\cal C}_9^{\perp}]=0$, relevant for the zero-point of the
FB asymmetry in the decay $B \to K^*\ell^+\ell^-$, we have factored in the
scale-dependence of $\zeta_\perp(\mu,q^2)$.
We do  a similar numerical analysis of $q_0^2$ in the PFF-scheme,
where the corresponding form factor $\zeta^P_\perp(q^2)$ is scale-independent,
and incorporate the effect of the logarithmic resummation in both the
$B \to K^* \gamma$ and $B \to K^* \ell^+\ell^-$ decays.
Solving now the equation
 $Re[{\cal C}_9^{\perp P}]=0$,
 using the central value of the soft form factor
$\zeta^P_\perp(0)$ obtained from the analysis of the $B\to K^*
\gamma$ branching ratio: $\zeta^P_\perp(0)=0.28$, and with all the other parameters
fixed at their central values given in Table~II,
 we find that in the PFF-scheme $q_0^2=3.98 \pm 0.18~\mbox{GeV}^2$. Had we
 dropped the resummation effect, we would get $q_0^2=4.03 \pm 0.22~\mbox{GeV}^2$,
 where the scale uncertainty $\Delta(q_0^2)_{\rm scale}=\pm 0.22$ GeV$^2$,
 derived here in the PFF-scheme, is
 consistent with the number $\Delta(q_0^2)_{\rm scale}=\pm 0.25$  GeV$^2$
obtained in \cite{Beneke01}.  Therefore, we conclude that the
difference in the
estimates of the scale dependence of $q_0^2$ here and in
Ref.~\cite{Beneke01} is both due to the incorporation of the SCET logarithmic
resummation and the different (scheme-dependent)
definitions of the effective form factors for the SCET currents 
and the ones used by Beneke et al.~\cite{Beneke01}.
 Using the SCET form factors defined in Eq.~(\ref{SCETff}) in this paper,
 we find that the scale-related uncertainty
$\Delta(q_0^2)_{\rm scale}$ is reduced than in the PFF-scheme of
Beneke et al.~\cite{Beneke01}. One expects that such scheme-dependent
differences will
become less marked after incorporating the $O(\alpha_s^2)$ effects in the
decay distributions for $B \to K^* \ell^+\ell^-$. Our comparative analysis 
hints at rather large $O(\alpha_s^2)$ corrections to $q_0^2$ in the PFF-scheme
and a moderate correction in the SCET analysis carried out by us in this
paper. Since the value of $q_0^2$ offers a precision test of the SM, and by
that token provides a window on the possible beyond-the-SM physics effects, it is
mandatory to undertake an  $O(\alpha_s^2)$ improvement of the current
theory of $B \to K^* \ell^+\ell^-$ decay.
As power corrections in
$1/m_b$ have not been considered here, although they are probably
comparable to the $O(\alpha_s)$ corrections as argued in a model-dependent
estimate of the $1/m_b$ corrections by Beneke et al.~\cite{Beneke01},
it also remains to be seen how a model-independent calculation of the same
effect the numerical value of $q_0^2$.  

\section{Summary}
In this paper, we have examined the rare B decay channel $B \to K^\ast \ell^+ \ell^-$
in the framework of SCET, where the factorization formula holds to all orders in
$\alpha_s$ and leading order in $1/m_b$. Making use of the existing
literature, we work with the relevant effective operators
in SCET and the corresponding matching procedures are discussed
in detail. The logarithms related to the different scales
$\mu_h=m_b$ and $\mu_l=\sqrt{m_b \Lambda_h}$ are  resummed by solving
numerically the renormalization group equation in SCET. We then give explicit
expressions for the differential distributions in $q^2$ for the
decay $B \to K^\ast \ell^+ \ell^-$ including the
 $O(\alpha_s)$ corrections. In the phenomenological analysis, we first
discuss the input parameters, especially how to extract the soft form factors
$\zeta_{\perp,\parallel}(q^2)$ from the full QCD form factors and also the constraints
on $\zeta_{\perp}(0)$ from the experimental data on the $B \to K^\ast \gamma$
decay.
Using the $q^2$-dependence of the form factors from the LCSRs and the
normalization
$\zeta_{\perp}(0)=0.32 \pm 0.02$ and $ \zeta_{\parallel}(0)=0.40 \pm 0.05$,
we work out the differential branching ratio and the forward-backward asymmetry as a
function of the dilepton invariant mass. In the region $1~ \mbox{GeV}^2 \le
q^2 \le 7~\mbox{GeV}^2$, where the perturbative method should be reliable, our
analysis yields
\begin{equation}
\int \limits_{1\mbox{\scriptsize ~GeV}^2}^{7\mbox{\scriptsize ~GeV}^2} d q^2
\frac{d Br(B^+ \to K^{\ast +} \ell^+ \ell^-)}{dq^2}=(2.92^{+0.67}_{-0.61}
)\times 10^{-7}~,
\end{equation}
which can be compared with the B factory measurements in the near future.
The largest uncertainty in the branching ratio is due to the imprecise
knowledge of $\zeta_{\parallel} (q^2)$. We have illustrated this by
using a value $\zeta_{\parallel} (0)=0.32$, which reduces the central value of
the branching ratio to $2.11 \times 10^{-7}$.
 We point out  that precisely measured
$q^2$-distributions in $B \to K^* \ell^+\ell^-$ and $B \to \rho \ell \nu_\ell$
would greatly reduce the form-factor related uncertainties in the differential
branching ratios. The FBA is less dependent on the soft form
factors, and the residual parametric dependencies are worked out.
We estimate the zero-point of the FBA to be
$q^2_0=(4.07^{+0.16}_{-0.13})~ \mbox{GeV}^2$. The stability of this result
against $O(\alpha_s^2)$ and $1/m_b$ corrections should be investigated in the
future.
\section*{Acknowledgement}
G.Z acknowledges the financial support from Alexander-von-Humboldt
Stiftung. G.Z is also grateful to D.s. Yang for useful discussions.
We thank Martin Beneke, Thorsten Feldmann, Alexander Parkhomenko and Dan
Pirjol for their comments on the earlier version of this manuscript and 
helpful communications.

\end{document}